\documentclass[%
reprint,
amsmath,amssymb,
aps,
]{revtex4-2}

\usepackage{graphicx}
\usepackage{dcolumn}
\usepackage{bm}
\usepackage{hyperref}

\usepackage{physics}
\usepackage{mathrsfs}
\usepackage{amsmath}
\usepackage{amssymb}
\usepackage{tikz}
\usepackage{subfigure} 
\allowdisplaybreaks

\usepackage{xcolor}
\usetikzlibrary{calc,decorations.markings,decorations.pathmorphing,shapes.misc}
\tikzset{
	branch point/.style={cross out,draw=black,fill=none,minimum size=2*(#1-\pgflinewidth),inner sep=0pt,outer sep=0pt}, 
	branch point/.default=5
}

\begin{document}
	
	
	\title{Relativistic free motion time of arrival operator for massive spin-0 particles with positive energy}
	
	\author{Philip Caesar Flores}
	\email{pmflores2@up.edu.ph}
	\author{Eric A. Galapon}%
	\email{eagalapon@up.edu.ph}
	\affiliation{%
		Theoretical Physics Group, National Institute of Physics \\ 
		University of the Philippines Diliman, 1101 Quezon City, Philippines
	}%
	
	%
	%
	
	\date{\today}
	
	\begin{abstract}
		A relativistic version of the Aharonov-Bohm time of arrival operator for spin-0 particles  was constructed by Razavi in [Il Nuovo Cimento B \textbf{63}, 271 (1969)]. We study the operator in detail by taking its rigged Hilbert space extension. It is shown that the rigged Hilbert space extension of the operator provides more insights into the time of arrival problem that goes beyond Razavi's original results. This allows us to use time of arrival eigenfunctions that exhibit unitary arrival to construct time of arrival distributions. The expectation value is also calculated and shown that particles can arrive earlier or later than expected classically. Lastly, the constructed time of arrival distribution, and expectation value are shown to be consistent with special relativity.   
	\end{abstract}
	
	\maketitle
	
	
	\section{Introduction}
	\label{sec:intro}
	
	At two extremes, classical mechanics branches out to relativity and quantum theory. However, the unification of the two theories under a single framework is hindered by the incompatible notions of time \cite{anderson2012problem}. Quantum mechanics treats time only as a parameter that governs the evolution of the system in the same way that we use time in Newton's equations of motion to calculate the position of a particle after a specific interval has elapsed. Meanwhile, time in general relativity plays a dynamic and intrinsic role in the evolution of the system. 
	
	The problem becomes apparent when we consider the time of arrival (TOA) of a particle. The classical TOA can be calculated by inverting the particle's equation of motion. In special relativity, we can calculate the TOA of a particle as measured in a particular Lorentz frame and perform a Lorentz transform to calculate the TOA as measured in another frame. However, standard quantum mechanics offers no solution to the TOA problem. 
	
	The prevalent von Neumann formulation of quantum mechanics implies that time is not an observable, thus, TOA is also not treated as an observable and it is non-sensical to ask the TOA of a quantum particle. The demotion of time as a parameter in quantum mechanics is mainly due to Pauli's theorem which states that there is no self-adjoint time operator that is canonically conjugate with its corresponding semi-bounded system Hamiltonian \cite{pauli1933handbuch}. This has lead the construction of TOA operators to either give up self-adjointness or conjugacy with the Hamiltonian to bypass Pauli's theorem. 
	
	The earliest detailed investigation on a TOA operator was due to Aharonov and Bohm \cite{aharonov1961time} where they considered a symmetric quantization of the classical TOA of a free particle 
	\begin{equation}
		\mathsf{\hat{T}_{AB}} = -\mu \dfrac{\mathsf{\hat{q}}\mathsf{\hat{p}}^{-1} + \mathsf{\hat{p}}^{-1} \mathsf{\hat{q}}}{2}, 
	\end{equation}
	where $\mu$ is the mass of the particle. The Aharonov-Bohm TOA operator is Hermitian but not self-adjoint, instead, it is a maximally symmetric operator \cite{muga1998space}. Grot, Rovelli, and Tate traces the non-self-adjointness of $\mathsf{\hat{T}_{AB}}$ to the singularity at $p=0$, and constructed a regularized TOA operator \cite{grot1996time}. A self-adjoint variant of $\mathsf{\hat{T}_{AB}}$ was proposed by Kijowski, Delagado, and Muga by considering the combination 
	\begin{equation}
		\mathsf{\hat{T}_{KDM}} = \mathsf{\hat{T}_{AB}}\Theta(\mathsf{\hat{p}}) - \mathsf{\hat{T}_{AB}}\Theta(-\mathsf{\hat{p}}),
	\end{equation}
	where $\Theta(z)$ is the heaviside function \cite{kijowski1974time,delgado1997arrival}. The first (second) term of $\mathsf{\hat{T}_{KDM}}$ only acts on the positive (negative) subspace of the system Hilbert space. This combination bypasses Pauli's theorem because $\mathsf{\hat{T}_{KDM}}$ is not canonically conjugate to the system Hamiltonian, instead, 
	\begin{equation}
		\comm{\text{sgn}(\mathsf{\hat{p}})\mathsf{\hat{H}}}{\mathsf{\hat{T}_{KDM}}} = i\hbar,
	\end{equation} 
	where $\text{sgn}(\mathsf{\hat{p}})$ is the sign function. 
	
	Pauli's proof was only formal, i.e. without regard to the domains of the operators involved and to the validity of the operations leading to his conclusion. One of us has shown that Pauli has made some implicit assumptions and that these were inconsistent \cite{galapon2002pauli}. This opens up an avenue to still consider TOA as an observable in standard quantum mechanics \cite{galapon2006theory,galapon2004shouldn,galapon2009theory,galapon2008quantum,galapon2004confined,galapon2005confined,sombillo2012quantum,galapon2005transition,muga2000arrival,egusquiza1999free,galapon2009quantum,flores2016synchronizing,flores2019quantum,galapon2018quantizations,pablico2020quantum,galapon2012only,dias2017space,ximenes2018comparing}. Moreover, it was discussed in Ref. \cite{galapon2009theory} that a time-operator based theory of quantum arrival has an unexpected connection with the collapse of the wavefunction on the appearance of a particle. In our pursuit to promote time as an observable in quantum mechanics, it is reasonable to investigate the possibility of a relativistic TOA operator to extend the insights of Ref. \cite{galapon2009theory} to the relativistic regime. 
	
	The construction of such an operator seems natural because special relativity tells us to treat space and time on equal footing as components of a four vector, thus it is natural to promote time as an observable with a corresponding operator because we also treat position as an operator. A caveat is that relativistic quantum mechanics is not a well-defined one-particle theory. This can lead to spontaneous pair creation such that if we perform a TOA experiment for a relativistic particle, then we are not sure if the particle that arrived is the same particle we started with. Nevertheless, we know that if one performs a TOA experiment, then we will obtain a TOA distribution. Assuming that TOA is a dynamical observable, is this TOA distribution generated by the spectral resolution of a corresponding TOA operator? If so, how do we construct this relativistic TOA operator?
	
	In non-relativistic quantum mechanics, one of us has proposed several methods to construct TOA operators for the free and interacting case by either solving the time-energy canonical commutation relation, quantizing the classical TOA, or by solving the time kernel equation \cite{galapon2018quantizations,galapon2004shouldn}. The resulting TOA operators are Hermitian, canonically conjugate with the system Hamiltonian, and reduces to the classical TOA in the limit as $\hbar\rightarrow0$. Furtheremore, the TOA eigenfunctions, which are referred as non-nodal and nodal, exhibit unitary arrival, i.e. they unitarily evolve through time to localize at the intended arrival point at their corresponding eigenvalues \cite{galapon2004confined,galapon2005confined,galapon2018quantizations,sombillo2016particle}. We postulate that the same properties should also be seen in our relativistic TOA operators. 
	
	In general, not all operators in non-relativistic quantum mechanics can be transferred to the relativistic regime because the eigenvalues and expectation values of the operator are not equal to the corresponding classical quantity \cite{greiner2000relativistic}. But despite the debates on time's status as a dynamical observable in non-relativistic quantum mechanics, e.g. time of arrival, there have been attempts to construct a corresponding relativistic TOA operator. The earliest construction of a such an operator was done by Razavi wherein TOA was treated as a constant of motion and applying a simple symmetric ordering to the quantization \cite{razavy1969quantum}. For spin-0 particles with positive energy $E_p=\sqrt{p^2c^2+\mu^2c^4}$, the TOA operator has the form 
	\begin{equation}
		\mathsf{\hat{T}_{Ra}} = - \dfrac{\mu}{2} \left( \sqrt{1 + \dfrac{\mathsf{\hat{p}}^2}{\mu^2c^2}} \mathsf{\hat{p}}^{-1}\mathsf{\hat{q}} + \mathsf{\hat{q}}\sqrt{1 + \dfrac{\mathsf{\hat{p}}^2}{\mu^2c^2}} \mathsf{\hat{p}}^{-1}\right),
		\label{eq:razaviopr}
	\end{equation}
	where, $\mu$ is the rest mass of the particle, and arrival point at x=0. This operator is canonically conjugate to the free Hamiltonian  $\sqrt{\mathsf{\hat{p}}^2 c^2 + \mu^2 c^4}$, and is the relativistic version of $\mathsf{\hat{T}_{AB}}$. For spin-1/2 particles, Razavi applied a Foldy-Wouthuysen transform on Eq. \eqref{eq:razaviopr}. The same operator for spin-0 particles was also recently obtained in Ref. \cite{Brunetti2010} by enlarging the system Hilbert space to include time as a dynamical observable. They then used the positive energy solution of the Klein-Gordon equation to construct a POVM whose first moment coincides with $\mathsf{\hat{T}_{Ra}}$.
	
	In spite of $\mathsf{\hat{T}_{Ra}}$ being the earlieast known relativistic free TOA operator, it was not studied in much detail in the same way as $\mathsf{\hat{T}_{AB}}$. There have been several other studies on the construction of relativsitic free TOA operators for spin-0 and spin-1/2 particles with each having proposed a different operator \cite{razavy1969quantum,leon1997time,wang2007relativistic,bauer2014dynamical,bauer2019time,aguillon2020time,bauer2020conditional,bunao2015one,bunao2015relativistic}. Another TOA operator for spin-0 particles was proposed in Ref. \cite{leon1997time} by inverting the equation of motion of the position operator using a suitable ordering rule that reduces to the non-relativistic TOA operator proposed by Grot, Rovelli, and Tate \cite{grot1996time}. A relativistic TOA operator for spin-1/2 particles was proposed in Ref. \cite{wang2007relativistic} using a total symmetrization of the relativistic TOA, and imposing conjugacy with the Dirac Hamiltonian in 1 dimension. Ref. \cite{bauer2014dynamical} proposed a self-adjoint TOA operator for spin-1/2 particles wherein the resulting commutation relation is analogous to the position-momentum commutation relation. Lastly, one-particle TOA operators for spin-0 and spin-1/2 particles were proposed in Refs. \cite{bunao2015one} and \cite{bunao2015relativistic} by solving the canonical commutation relation using the Bender-Dunne basis operators \cite{bender1989exact,bender1989integration,bender2013advanced}, and performing a Feshbach-Villars transformation \cite{feshbach1958elementary,greiner2000relativistic} to diagonalize the TOA operator which separates the positive and negative energy components of the wavefunctions. Explicitly, the action of the one-particle operator of Refs. \cite{bunao2015one} and \cite{bunao2015relativistic} on a wavefunction $\tilde{\psi}(p)$ is 
	\begin{align}
		\mathsf{\hat{T}_{BG}} \tilde{\psi}(p) =& -\hat{\sigma}_3 \dfrac{i\hbar}{c^2} \left( \dfrac{E_p}{p} \dfrac{d}{dp} - \dfrac{1}{2} \dfrac{\mu^2c^4}{p^2 E_p} \right)\tilde{\psi}(p) \nonumber \\
		=&\hat{\sigma}_3 \mathsf{\hat{T}_{Ra}}\tilde{\psi}(p)
	\end{align} 
	where, $\hat{\sigma}_3$ is a Pauli spin matrix. This implies that $\mathsf{\hat{T}_{Ra}}$ is the positive energy component of the one-particle operator $\mathsf{\hat{T}_{BG}}$.
	
	The purpose of this paper is to provide a detailed study of the operator $\mathsf{\hat{T}_{Ra}}$ by taking its position space representation within the rigged Hilbert space (RHS) formulation of quantum mechanics \cite{galapon2018quantizations,Madrid2002a,Madrid2002,Madrid2003,bohm1974rigged}. This is in contrast to Razavi's results \cite{razavy1969quantum} wherein he used the momentum space representation of the operator. It turns out that the rigged Hilbert space extension of  $\mathsf{\hat{T}_{Ra}}$ provides more insight to the TOA problem than that of Razavi's. Specifically, its eigenfunctions exhibit unitary arrival in the same way as that of the non-relativistic TOA operators in Ref. \cite{galapon2018quantizations}. This allows us to provide a meaningful interpretation to the operator $\mathsf{\hat{T}_{Ra}}$ that is beyond Razavi's results. 
	
	The rest of the paper is structured as follows. Sec. \ref{sec:quant-freetoa-opr} discusses the construction of the rigged Hilbert space extension of $\mathsf{\hat{T}_{Ra}}$. The corresponding eigenvalue equation is then solved numerically in Sec. \ref{sec:eigenfunctions} to investigate the dynamics of the eigenfunctions, which are then compared with the ones solved analytically by Razavi in Ref. \cite{razavy1969quantum}. It is also shown how Razavi's results can be modified to recover the non-nodal and nodal eigenfunctions. In Sec. \ref{sec:expecval}, we calculate the expected TOA of a single-peaked wavepacket and show that there are quantum correction terms to the relativistic TOA of a quantum particle. The extent of these quantum correction terms is demonstrated in Sec. \ref{sec:quantcorr} by considering a Gaussian wavepacket. The corresponding TOA distributions are then constructed in Sec. \ref{sec:TOAdist}, and shown to spread to \lq\lq superluminal\rq\rq times of arrival due to the non-locality of the wavepacket. Furthermore, TOA distributions are constructed to demonstrate that the Hamiltonian is a generator of time translation. Lastly, Sec. \ref{sec:conc} summarizes the paper.

	\section{Quantized relativistic free time of arrival operator}
	\label{sec:quant-freetoa-opr}
	
	We describe a quantum particle within the RHS formulation of quantum mechanics \cite{galapon2018quantizations,Madrid2002a,Madrid2002,Madrid2003,bohm1974rigged} to accomodate non-square integrable functions that are outside the Hilbert space, e.g. Dirac-delta function (plane wave) which is an eigenfunction of the position (momentum) operator. In our case, we choose the fundamental space of our RHS to be the  the space of infinitely continuously differentiable complex valued functions with compact supports $\Phi$ such that the RHS is $\Phi\subset L^2(\mathbb{R}) \subset \Phi^\times$. The standard Hilbert space formulation of quantum mechanics is recovered by taking the closures on $\Phi$ with respect to the metric of $L^2(\mathbb{R})$. 
	
	In coordinate representation, a quantum observable $\mathsf{\hat{A}}$ is a mapping from $\Phi$ to $\Phi^\times$, and is given by the formal integral operator 
	\begin{equation}
		(\mathsf{\hat{A}}\varphi)(q) = \int_{-\infty}^\infty dq' \mel{q}{\mathsf{\hat{A}}}{q'} \varphi(q')
		\label{eq:genintop}
	\end{equation}
	where the kernel satisfies $ \mel{q}{\mathsf{\hat{A}}}{q'}= \mel{q'}{\mathsf{\hat{A}}}{q}^*$, to ensure Hermiticity. In general, the integral Eq. \eqref{eq:genintop} is interpreted in the distributional sense, i.e. it is a functional on $\Phi$ wherein the kernel is a distribution. By doing so, we are able to treat observables as scalar objects. In our present case, we will see that $\mathsf{\hat{T}_{Ra}}$ becomes an integral operator with a regular kernel.
	
	It follows that the position space representation of $\mathsf{\hat{T}_{Ra}}$ is 
	\begin{equation}
		(\mathsf{\hat{T}_{Ra}}\varphi)(q) = \int_{-\infty}^\infty dq' \mel{q}{\mathsf{\hat{T}_{Ra}}}{q'} \varphi(q')
		\label{eq:toaoprint}
	\end{equation}  
	wherein the time kernel is given by  
	\begin{align}
		\mel{q}{\mathsf{\hat{T}_{Ra}}}{q'} = -\mu \dfrac{q+q'}{2} \mel{q}{\mathsf{\hat{p}}^{-1} \sqrt{1 + \dfrac{\mathsf{\hat{p}}^2}{\mu^2 c^2}}}{q'}.
		\label{eq:pkernelfree}
	\end{align}
	The right hand side of Eq. \eqref{eq:pkernelfree} is evaluated by inserting the resolution of the identity $\mathsf{1}=\int_{-\infty}^{\infty} dp \ket{p}\bra{p}$, and using the plane wave expansion $\braket{q}{p}=e^{iqp/\hbar}/\sqrt{2\pi\hbar}$, which yields the expression 
	\begin{align}
		\langle q|& \mathsf{\hat{p}}^{-1} \sqrt{1 + \dfrac{\mathsf{\hat{p}}^2}{\mu^2 c^2}}  | q' \rangle \nonumber \\
		=& \int_{-\infty}^{\infty} \frac{dp}{2\pi\hbar}  \exp[\frac{i}{\hbar}(q-q')p] \frac{1}{p}  \left(  \sqrt{1 + \frac{p^2}{\mu^2 c^2}} \right).
		\label{eq:pkernel0}
	\end{align}
	The integral Eq. \eqref{eq:pkernel0} is equal to the Cauchy principal value and is evaluated using the method in Ref. \cite{Galapon2016}, i.e.
	\begin{align}
		\langle q|& \mathsf{\hat{p}}^{-1} \sqrt{1 + \dfrac{\mathsf{\hat{p}}^2}{\mu^2 c^2}}  | q' \rangle \nonumber \\
		=& \dfrac{i}{2\hbar} \left( 1 + \dfrac{2}{\pi} \int_1^\infty dz \exp(-\dfrac{\mu c}{\hbar} \abs{q-q'} z) \dfrac{\sqrt{z^2-1}}{z} \right) \nonumber \\
		&\times \text{sgn}(q-q').
		\label{eq:pkernel1}
	\end{align} 
	See Appendix \ref{sec:pkernel} for details. Thus, the time kernel is now 
	\begin{align}
		\mel{q}{\mathsf{\hat{T}_{Ra}}}{q'} =&  \dfrac{\mu}{i\hbar} \left( \dfrac{q+q'}{4}\right) T_c(q,q') \text{sgn}(q-q')
		\label{eq:timekernel}
	\end{align}
	where,
	\begin{align}
		T_c(q,q') =& 1 + \dfrac{2}{\pi} \int_1^\infty dz \exp(-\dfrac{\mu c}{\hbar} \abs{q-q'} z) \dfrac{\sqrt{z^2-1}}{z}.
		\label{eq:rel_freekernel}
	\end{align}	
	It is easy to see that $\lim_{c\rightarrow\infty}T_c(q,q')=1$ since the integral term vanishes, which reduces Eq. \eqref{eq:timekernel} to the known kernel of the Aharonov-Bohm TOA operator \cite{galapon2018quantizations}, i.e.
	\begin{align}
		\lim_{c\rightarrow\infty}\mel{q}{\mathsf{\hat{T}_{Ra}}}{q'} = &\mel{q}{\mathsf{\hat{T}_{AB}}}{q'} = \dfrac{\mu}{i\hbar} \left( \dfrac{q+q'}{4}\right) \text{sgn}(q-q').
	\end{align}
	
	For completeness, a closed form expression for the integral term in Eq. \eqref{eq:rel_freekernel} can be obtained using the substitution $z=\sec\theta$ and the identiy
	\begin{equation}
		K_1(a) = \int_0^\frac{\pi}{2} d\theta \sec^2\theta e^{-a\sec\theta}
	\end{equation}
	where $K_n(a)$ is the modified bessel function of the second kind. This will make
	\begin{align}
		T_c(q,q') =& \dfrac{2}{\pi} K_1\left( \dfrac{\mu c}{\hbar} \abs{q-q'} \right) \nonumber \\
		&+ \dfrac{\mu c}{\hbar} \abs{q-q'} K_0\left( \dfrac{\mu c}{\hbar} \abs{q-q'} \right) L_{-1}\left( \dfrac{\mu c}{\hbar} \abs{q-q'} \right) \nonumber \\
		&+ \dfrac{\mu c}{\hbar} \abs{q-q'} K_1\left( \dfrac{\mu c}{\hbar} \abs{q-q'} \right) L_0\left( \dfrac{\mu c}{\hbar} \abs{q-q'} \right)
	\end{align}
	where $L_n(z)$ is the modified Struve function. For practical purposes, we will use the integral form Eq. \eqref{eq:rel_freekernel} in the next sections because it separates the relativistic and non-relativistic terms of the TOA.

	\section{Dynamics of the TOA eigenfunctions in position space}
	\label{sec:eigenfunctions}
	
	A time operator $\mathsf{\hat{T}}$ is a legitimate TOA operator if the eigenfunctions exhibit unitary arrival, i.e. they unitarily evolve through time to localize at the intended arrival point at their corresponding eigenvalues \cite{galapon2004confined,galapon2005confined,galapon2018quantizations,sombillo2016particle}. In this section, we investigate the eigenfunctions of the rigged Hilbert space extension of  $\mathsf{\hat{T}_{Ra}}$ given by Eq. \eqref{eq:toaoprint}, and their dynamics. We compare these with the eigenfunctions solved by Razavi \cite{razavy1969quantum}.

	\subsection{Eigenfunctions of the time kernel}
	\label{subsec:eig_kernel}
	
	To study, the dynamics of the TOA eigenfunctions of the kernel $\mel{q}{\mathsf{\hat{T}_{Ra}}}{q'}$, we need to solve the corresponding eigenvalue equation. It follows from Eq. \eqref{eq:toaoprint} and \eqref{eq:timekernel} that the relevant eigenvalue problem is 
	\begin{align}
		\int_{-\infty}^\infty & dq' \left[ 1 + \dfrac{2}{\pi} \int_1^\infty dz \exp(-\dfrac{\mu c}{\hbar} \abs{q-q'} z) \dfrac{\sqrt{z^2-1}}{z} \right]   \nonumber \\
		& \times \dfrac{\mu}{i\hbar} \left( \dfrac{q+q'}{4}\right)  \text{sgn}(q-q') \tilde{\Phi}_\tau(q') \nonumber \\
		=&\tau \tilde{\Phi}_\tau(q),
		\label{eq:inteig}
	\end{align}
	where, $\tilde{\Phi}_\tau(q)$ is the TOA eigenfunction with an eigenvalue $\tau$. However, solving this integral equation is intractable and we must proceed to a numerical solution by coarse-graining. This is done by confining the system in a box centered at the origin with length $2l$ to project the operator in the Hilbert space $\mathcal{H}_l=L^2[-l,l]$. The eigenvalue problem is then solved by quadrature using the Nystrom method and the dynamics of the eigenfunctions were constructed by modifying the methods outlined in Refs. \cite{galapon2004confined,galapon2006theory,galapon2018quantizations} to allow evolution via the Klein-Gordon equation (see Appendix \ref{sec:coarsegrain} for details). The behavior as $l\rightarrow\infty$ is investigated by successivley increasing the confining length. 
	
	\begin{figure}[t!]		
		\centering
		\includegraphics[width=0.45\textwidth]{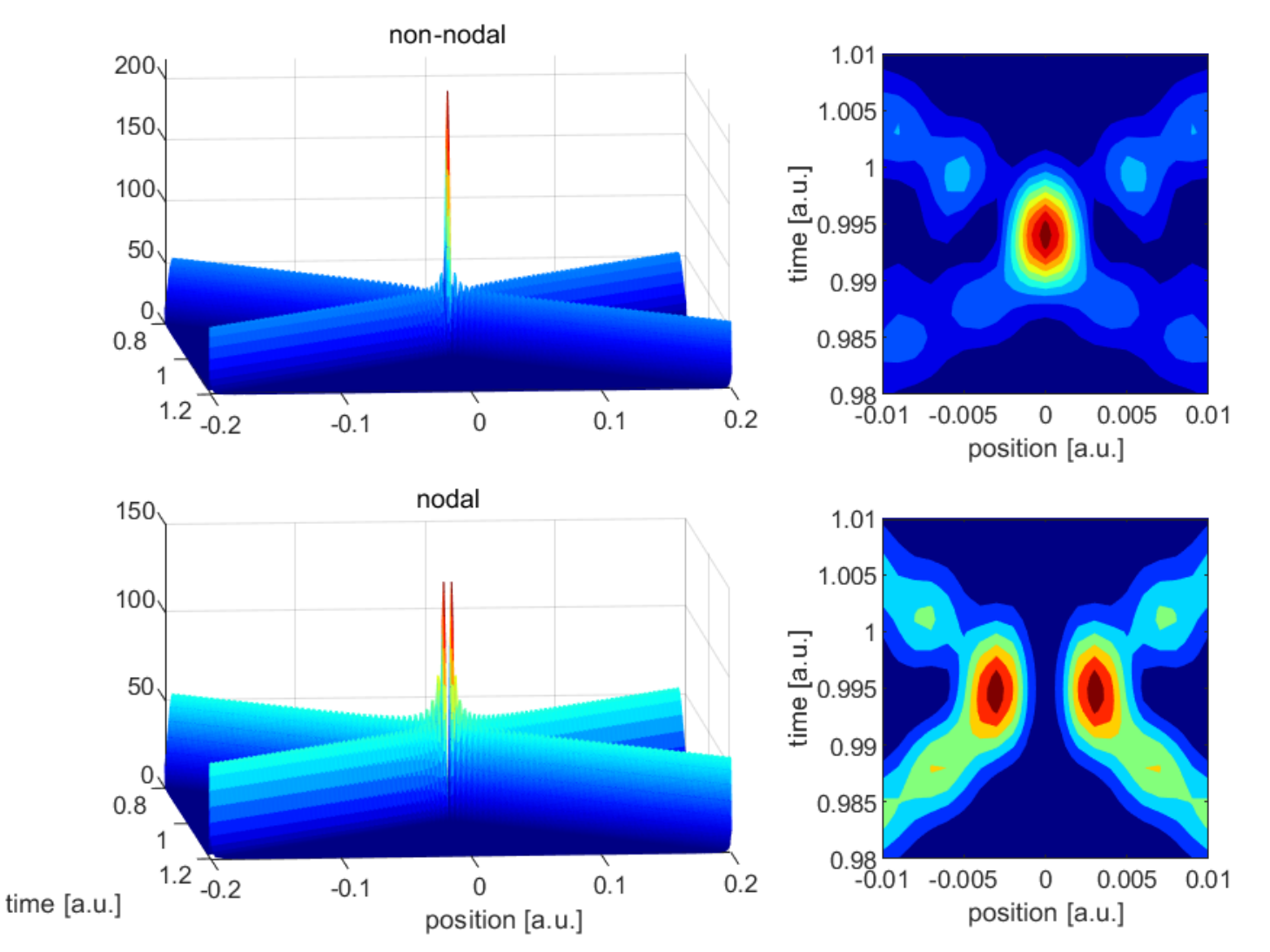}
		\caption{Eigenfunction dynamics of the time kernel $\mel{q}{\mathsf{\hat{T}_{Ra}}}{q'}$ confined in the region $[-1,1]$ with eigenvalue $\tau=0.9944$ for the parameters $\mu=c=\hbar=1$.}
		\label{fig:eigenevo-integ}
	\end{figure}
	
	The eigenfunctions of Eq. \eqref{eq:toaoprint} are two-fold degenerate and are categorized into non-nodal and nodal eigenfunctions whose dynamics are shown in Fig. \ref{fig:eigenevo-integ}. The former has a single peak that gathers at the arrival point with its minimum width occuring at its eigenvalue, while the latter has two peaks that gather at the arrival point with their closest separation occuring at its eigenvalue. This behavior is also present for all the non-relativistic quantized TOA operators  \cite{galapon2004confined,galapon2005confined,galapon2018quantizations}. We thus claim that RHS extension of $\mathsf{\hat{T}_{Ra}}$ represents a legitimate TOA operator.
	
	\begin{figure}[t]
		\centering
		\includegraphics[width=0.45\textwidth]{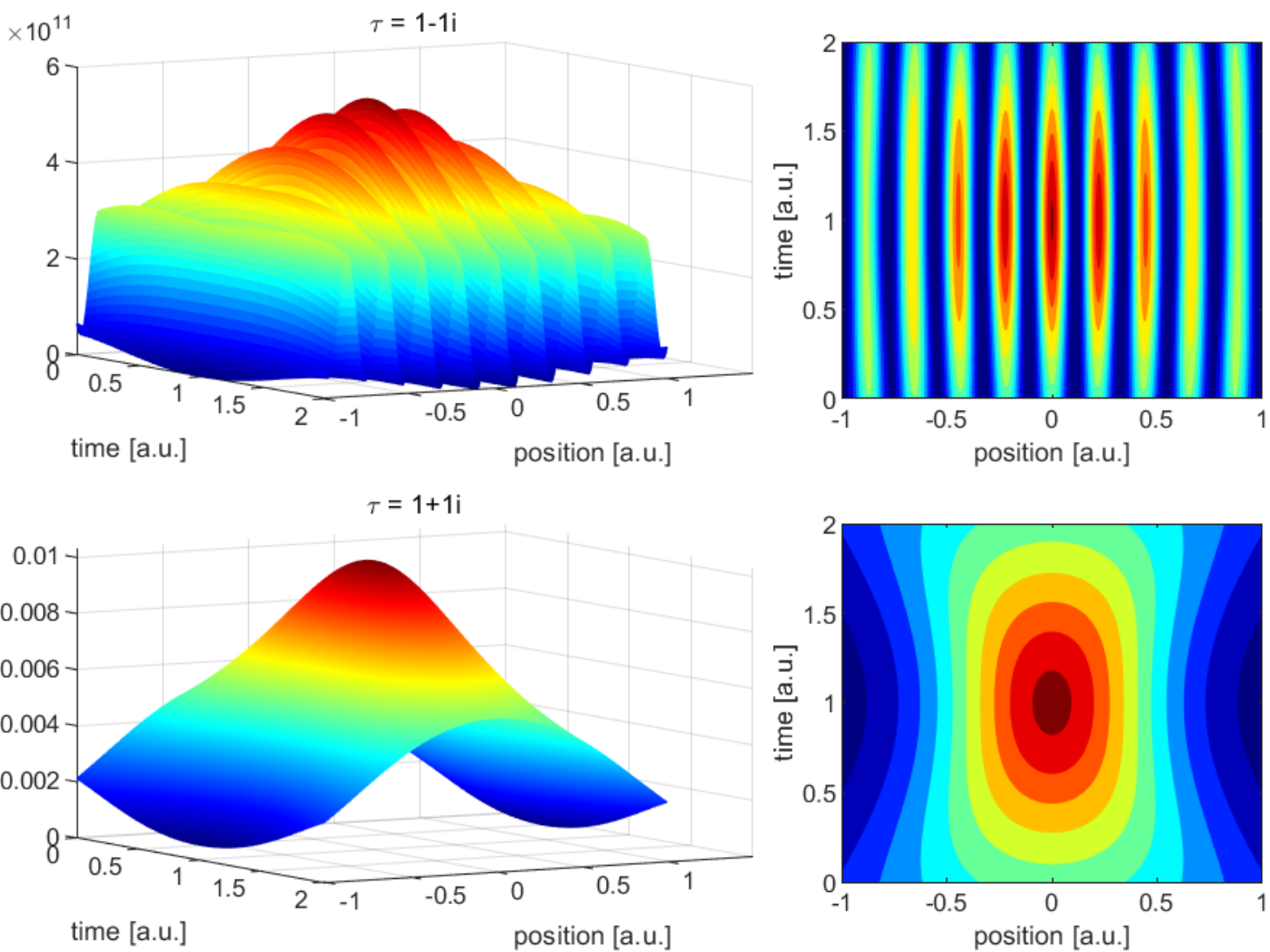}
		\caption{Dynamics of Razavi's unnormalized TOA eigenfunctions Eq. \eqref{eq:eigs_complex} with complex-eigenvalue $\tau=1\mp1i$ for the parameters $\mu=c=\hbar=1$ and converging factor $e^{-0.001 p^2}$.}
		\label{fig:eigenevo-razavi-comp}
	\end{figure}
	\subsection{Razavi's TOA eigenfunctions}
	
	Razavi obtained an analytic form of the TOA eigenfunctions \cite{razavy1969quantum} by solving the eigenvalue equation in momentum representation, i.e. 
	\begin{align}
		\mathsf{\hat{T}_{Ra}} & \phi_{\tau}(p) \nonumber \\
		=& - \dfrac{\mu i \hbar}{2} \left( \sqrt{1 + \dfrac{p^2}{\mu^2c^2}} \dfrac{1}{p} \dfrac{\partial}{\partial p} + \dfrac{\partial}{\partial p}\sqrt{1 + \dfrac{p^2}{\mu^2c^2}} \dfrac{1}{p}\right) \phi_{\tau}(p) \nonumber \\
		=& - \dfrac{i\hbar}{c^2} \left[ \dfrac{E_p}{p} \dfrac{\partial}{\partial p} - \dfrac{1}{2} \dfrac{\mu^2 c^4}{p^2 E_p}\right] \phi_{\tau}(p) = \tau \phi_{\tau}(p).
		\label{eq:eigeq}
	\end{align}
	The eigenfunctions are then given as 
	\begin{align}
		\phi_{\tau}(p) = N \sqrt{\dfrac{|p|c}{E_p}} \exp(\dfrac{i}{\hbar}E_p\tau),
		\label{eq:eigs_complex}
	\end{align}
	where $N$ is a normalization constant. He then concluded that the eigenvalue $\tau$ must be complex-valued so that the eigenfunction is square integrable. For real-valued $\tau$, the eigenfunction is obtained from Eq. \eqref{eq:eigs_complex} by performing the integration
	\begin{align}
		\phi_{\tau}^{(Re)}(\epsilon,p) =& \int_{\tau-\epsilon}^{\tau+\epsilon} d\tau' \phi_{\tau'}(p) \nonumber \\
		=& \sqrt{\dfrac{\hbar}{\pi\epsilon}} \dfrac{\sin(\epsilon E_p/\hbar)}{E_p} \sqrt{\dfrac{|p|c}{E_p}} \exp(\dfrac{i}{\hbar}E_p\tau)
		\label{eq:eigs_real}
	\end{align}
	where, $\epsilon\rightarrow0$. The dynamics of the eigenfunctions Eq. \eqref{eq:eigs_complex} and \eqref{eq:eigs_real} in position space are constructed via 
	\begin{align}
		\tilde{\phi}_\tau(q,t) =& \int_{-\infty}^\infty \dfrac{dp}{\sqrt{2\pi\hbar}} e^{ipq/\hbar} e^{-iE_pt/\hbar} \phi_\tau(p).
		\label{eq:eigenevo}
	\end{align}
	In anticipation that the integral Eq. \eqref{eq:eigenevo} may diverge, we insert a converging factor $\lim_{\delta\rightarrow0}e^{-\delta p^2}$. 
	
	Fig. \ref{fig:eigenevo-razavi-comp} shows the dynamics of the unnormalized eigenfunctions $\phi_{\tau}(p)$ in Eq. \eqref{eq:eigs_complex} with eigenvalues $\tau_\pm=1\pm i$. It can be seen that the eigenfunctions do not exhibit unitary arrival compared to the non-nodal and nodal eigenfunction in Fig. \ref{fig:eigenevo-integ}. Specifically, for $\tau_{-}$ the probability density oscillates and there is no sharp localization at each peak. Moreover, for $\tau_{+}$ there is a single peak that gathers at the arrival point at a time equal to $\Re{\tau_{+}}=1$ but no sharp localization is observed. Thus, the operator $\mathsf{\hat{T}_{Ra}}$ is not a legitimate TOA operator when it is expressed as the first moment of the identity generated by the eigenfunctions Eq. \eqref{eq:eigs_complex}.
	
	\begin{figure}[t!]
		\centering
		\includegraphics[width=0.45\textwidth]{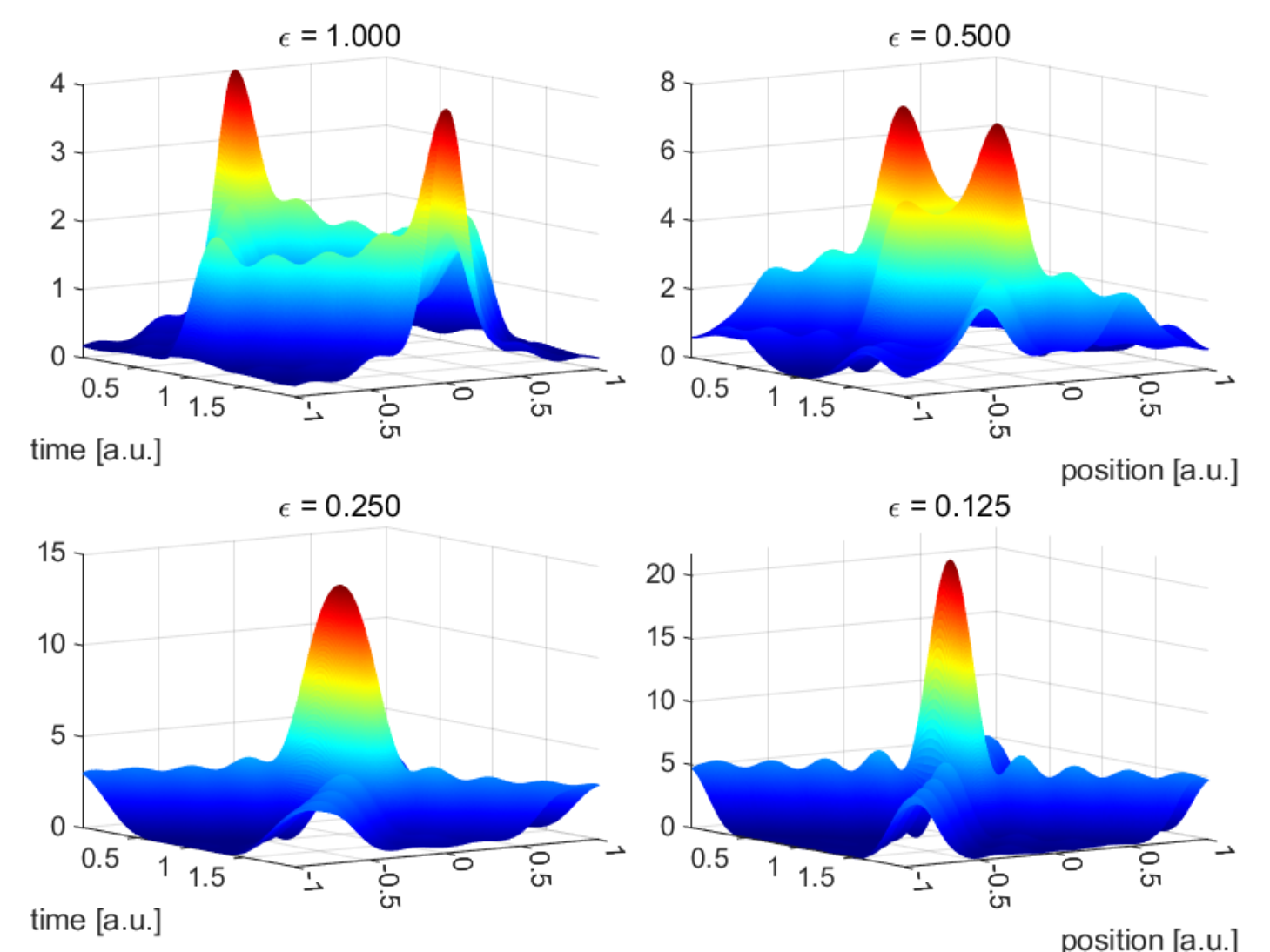}
		\caption{Dynamics of Razavi's TOA eigenfunctions Eq. \eqref{eq:eigs_real} with real-eigenvalue $\tau=1$ for the parameters $\mu=c=\hbar=1$. The parameter $\epsilon$ causes two peaks to gather that are separated in time. The peaks coincide as $\epsilon\rightarrow0$.}
		\label{fig:eigenevo-razavi-real}
	\end{figure}
	
	Fig. \ref{fig:eigenevo-razavi-real} shows the dynamics of the eigenfunctions $\phi_{\tau}^{(Re)}(\epsilon,p)$ in Eq. \eqref{eq:eigs_real} with eigenvalues $\tau=1$. It can be seen that as $\epsilon\rightarrow0$, the eigenfunction $\phi_{\tau}^{(Re)}(\epsilon,p)$ is able to capture the dynamics of the non-nodal eigenfunction in Fig. \ref{fig:eigenevo-integ} and exhibits unitary arrival. However, $\phi_{\tau}^{(Re)}(\epsilon,p)$ is not able to reproduce the dynamics of the nodal eigenfunction. As such, the operator $\mathsf{\hat{T}_{Ra}}$ will also not be a legitimate TOA operator when expressed as the first moment of the identity generated by the eigenfunctions Eq. \eqref{eq:eigs_real}.

	\subsection{Recovering the non-nodal and nodal eigenfunction from Razavi's eigenfunctions}
	
	The analytic form of the non-nodal and nodal eigenfunctions in momentum representation can be obtained by also solving Eq. \eqref{eq:eigeq}. This is done by treating the eigenfunction $\phi_\tau(p)$ as a distribution for the cases $p<0$ and $p>0$. Incidentally, the same problem was solved in Refs.  \cite{bunao2015one,bunao2015relativistic} which yields
	\begin{align}
		\phi_\tau^{(\pm)}(p) = \sqrt{\dfrac{c}{4\pi\hbar}} \sqrt{\dfrac{|p|c}{E_p}} \exp(\dfrac{i}{\hbar}E_p\tau) \Theta(\pm p). 
	\end{align}
	wherein the eigenvalue $\tau$ is real. The heaviside function $\Theta(\pm p)$ imposes the eigenfunctions $\phi_\tau^{(\pm)}(p)$ to have support on either positive or negative momentum only. 
	
	The non-nodal and nodal eigenfunctions, $\Phi_{\tau}^{(non)}(p)$ and $\Phi_{\tau}^{(nod)}(p)$, respectively, are then constructed by taking the sum and difference of $\phi_\tau^{(\pm)}(p)$, i.e. 
	\begin{align}
		\Phi_{\tau}^{(non)}(p) =& \phi_\tau^{(+)}(p) + \phi_\tau^{(-)}(p)\nonumber \\
		=& \sqrt{\dfrac{c}{4\pi\hbar}} \sqrt{\dfrac{|p|c}{E_p}} \exp(\dfrac{i}{\hbar}E_p\tau) \label{eq:noneig}\\ 
		\Phi_{\tau}^{(nod)}(p) =& \phi_\tau^{(+)}(p) - \phi_\tau^{(-)}(p) \nonumber \\
		=& \sqrt{\dfrac{c}{4\pi\hbar}} \sqrt{\dfrac{|p|c}{E_p}} \exp(\dfrac{i}{\hbar}E_p\tau) \text{sgn}(p)
		\label{eq:nodeig}
	\end{align} 
	Notice that the non-nodal eigenfunction Eq. \eqref{eq:noneig} and Razavi's eigenfunction Eq. \eqref{eq:eigs_complex} are equal if he concluded that the eigenvalues are real-valued. The non-nodal and nodal eigenfunctions Eqs. \eqref{eq:noneig}-\eqref{eq:nodeig} are complete but non-orthogonal \cite{bunao2015one,bunao2015relativistic}. This makes $\mathsf{\hat{T}_{Ra}}$ a maximally symmetric operator which is also true for $\mathsf{\hat{T}_{AB}}$ \cite{muga1998space,galapon2004confined,galapon2005confined,galapon2005transition,egusquiza1999free}. Furthermore, in the limit as $c\rightarrow\infty$, the eigenfunctions Eqs. \eqref{eq:noneig}-\eqref{eq:nodeig} reduces to the non-nodal and nodal eigenfunctions of $\mathsf{\hat{T}_{AB}}$ up to a phase factor \cite{bunao2015one,bunao2015relativistic}.   
	
	\begin{figure}[t!]
		\centering
		\includegraphics[width=0.45\textwidth]{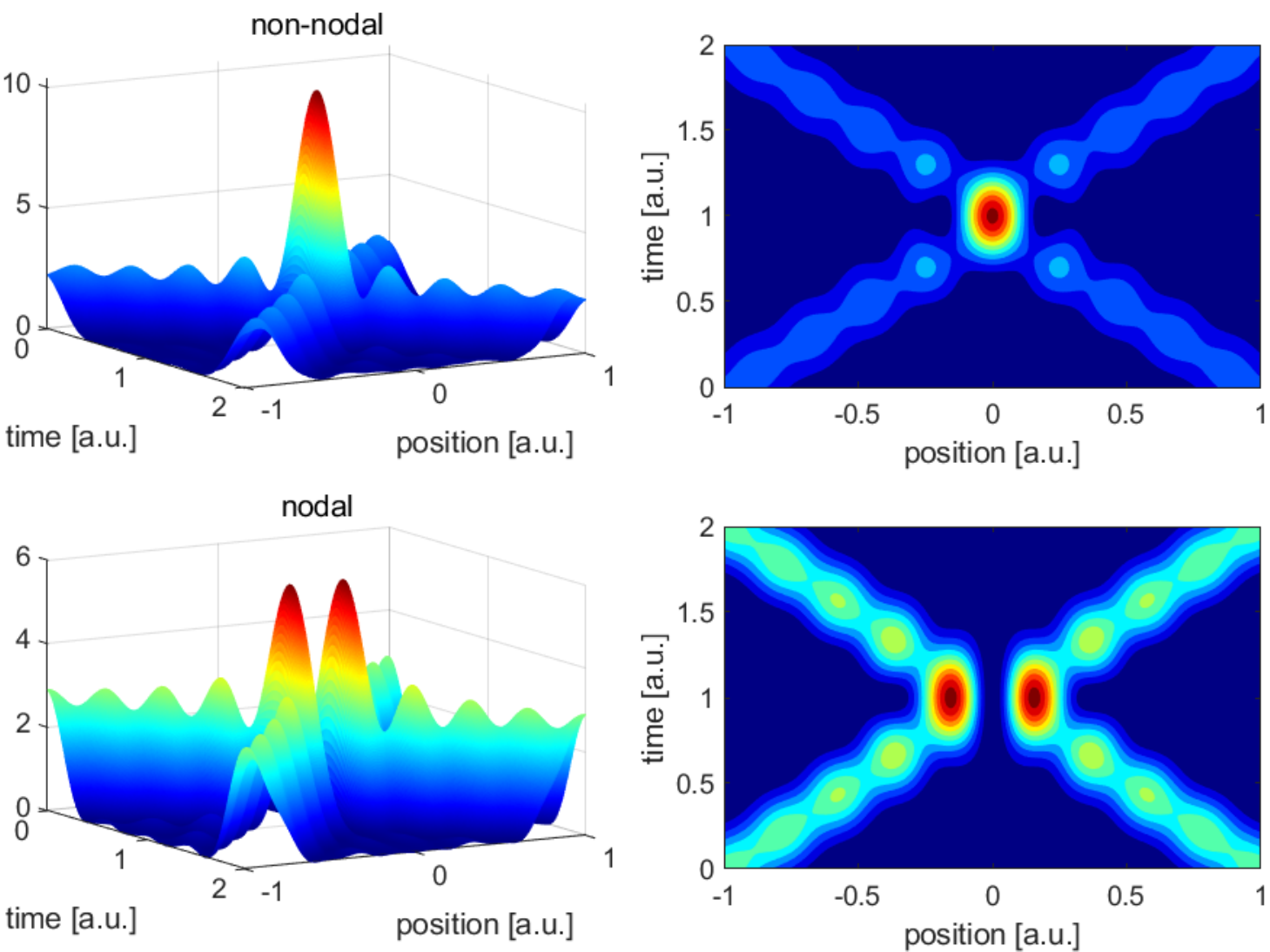}
		\caption{Dynamics of the analytic non-nodal and nodal eigenfunctions Eqs. \eqref{eq:noneig}-\eqref{eq:nodeig} for the relativistic free TOA operator with eigenvalue $\tau=1$ for the parameters $\mu=c=\hbar=1$ and converging factor $e^{-0.001 p^2}$.}
		\label{fig:eigenevo}
	\end{figure}
	
	The dynamics of the non-nodal and nodal eigenfunctions Eqs. \eqref{eq:noneig}-\eqref{eq:nodeig} in position space are then constructed using Eq. \eqref{eq:eigenevo} and inserting a converging factor $\lim_{\delta\rightarrow0}e^{-\delta p^2}$. Figure \ref{fig:eigenevo} shows a pair of evolving non-nodal and nodal eigenfunctions which captures the same dynamics as that of Fig. \ref{fig:eigenevo-integ} and exhibits unitary arrival. However, the localization is not as sharply defined as that of Fig. \ref{fig:eigenevo-integ} which is a consequence of coarse-graining the operator.

	\section{Expectation value of the relativistic free time of arrival operator}
	\label{sec:expecval}
	
	Let the expected time of arrival be the average of an arbitrarily large number of independent measurements of the TOA at the origin. We assume that the initial wavefunction is prepared in a pure state $\psi(q)=e^{ipq/\hbar}\varphi(q)$, wherein the wavepacket $\varphi(q)$ is independent of $\hbar$ and satifsies $\mel{\varphi}{\mathsf{\hat{p}}}{\varphi}=0$. The expected time of arrival is explicitly given as 
	\begin{align}
		\tau =& \int_{-\infty}^\infty\int_{-\infty}^\infty dq dq' \psi^*(q)\mel{q}{\mathsf{\hat{T}_{Ra}}}{q'}\psi(q')=\tau_0 + \tau_c
		\label{eq:tau}
	\end{align}
	where,
	\begin{align}
		\tau_0 =&\int_{-\infty}^{\infty} dq \varphi^*(q)e^{-ipq/\hbar} \nonumber \\
		&\times \int_{-\infty}^{\infty}  dq' e^{ipq'/\hbar} \varphi(q')  \left[ \dfrac{\mu}{4i\hbar} (q+q') \text{sgn}(q-q')\right] 
		\label{eq:tau0}
	\end{align}
	\begin{align}
		\tau_c =&\int_{-\infty}^{\infty} dq \varphi^*(q)e^{-ipq/\hbar}  \nonumber \\
		&\times  \int_{-\infty}^{\infty} dq' e^{ipq'/\hbar} \varphi(q')  \left[ \dfrac{\mu}{4i\hbar} (q+q') \text{sgn}(q-q')\right] \nonumber \\
		&\times \left[ \dfrac{2}{\pi} \int_{1}^{\infty} dz \dfrac{\sqrt{z^2-1}}{z} \exp(-\dfrac{\mu c}{\hbar} \abs{q-q'} z)\right]
		\label{eq:tauc}
	\end{align}

	It is easy to see that $\tau_0$ is just equal to the expectation value of the time of arrival of a non-relativistic free particle because it is independent of $c$, while $\tau_c$ is the relativistic correction. Now, $\tau_0$ was already calculated in Ref. \cite{galapon2009quantum} and is explicitly given as 
	\begin{align}
		\tau_0 =& -\mu \sum_{n=0}^{\infty} \dfrac{\hbar^{2n} \chi_1^{(n)} }{p^{2n+1}} + \mu  \sum_{n=0}^{\infty} \dfrac{(-1)^n \hbar^{2n+1} \chi_2^{(n)} }{p^{2n+2}}
	\end{align}
	where, 
	\begin{align}
		\chi_1^{(n)} =& \int_{-\infty}^{\infty} dq q \abs{\varphi^{(n)}(q)}^2 q \\
		\chi_2^{(n)} =& \int_{-\infty}^{\infty} dq \text{Im} \left[ \varphi^*(q)\varphi^{(2n+1)}(q)\right]q .
	\end{align}
	To evaluate $\tau_c$, it will be convenient to interchange the orders of integration as follows
	\begin{align}
		\tau_c = \dfrac{2}{\pi} \int_1^\infty dz \dfrac{\sqrt{z^2-1}}{z} \int_{-\infty}^{\infty} dq e^{-ipq/\hbar} \varphi^*(q) \psi_p(q)
		\label{eq:tauc_a}
	\end{align}
	where 
	\begin{align}
		\psi_p(q) =& e^{-\mu c q z / \hbar} \dfrac{\mu}{4i\hbar} \int_{-\infty}^{q} dq' (q+q') \varphi(q') e^{i (p-i\mu c z)q/\hbar} \nonumber \\
		&- e^{\mu c q z / \hbar} \dfrac{\mu}{4i\hbar} \int_{q}^{\infty} dq' (q+q') \varphi(q') e^{i (p+i\mu c z)q/\hbar}
		\label{eq:tau_psi}
	\end{align}
	We apply repeated integration by parts on Eq. \eqref{eq:tau_psi} and follow the steps used in Ref. \cite{galapon2009quantum} which leads to 
	\begin{align}
		\tau_c =& -\mu \sum_{n=0}^{\infty} \hbar^{2n} \Omega_{2n}(p) \chi_1^{(n)} \nonumber \\
		& + \mu  \sum_{n=0}^{\infty}(-1)^{n}\hbar^{2n+1} \Omega_{2n+1}(p) \chi_2^{(n)}
	\end{align}
	where
	\begin{equation}
		\Omega_n(p) = \dfrac{2}{\pi} \int_1^\infty dz \dfrac{\sqrt{z^2-1}}{z} \dfrac{\Re[(p+i\mu c z)^{n+1}]}{(p^2+\mu^2c^2z^2)^{n+1}}
	\end{equation}
	Combining the results for $\tau_0$ and $\tau_c$ we get
	\begin{align}
		\tau =& -\mu \sum_{n=0}^{\infty} \dfrac{\hbar^{2n} }{p^{2n+1}} \gamma_c^{(2n)} \chi_1^{(n)} \nonumber \\
		& + \mu  \sum_{n=0}^{\infty} \dfrac{(-1)^n \hbar^{2n+1}  }{p^{2n+2}} \gamma_c^{(2n+1)}\chi_2^{(n)}
		\label{eq:TOAexpec}
	\end{align}
	where
	\begin{align}
		\gamma_c^{(n)}(p) = 1 + \dfrac{2}{\pi} \int_1^\infty dz & \dfrac{\sqrt{z^2-1}}{z} \left( \dfrac{p}{p^2+\mu^2c^2z^2} \right)^{n+1} \nonumber \\
		&\times \text{Re}\left[(p+i\mu c z)^{n+1}\right]
		\label{eq:TOAexpecgamma}
	\end{align}	
	Generally, the series expansion Eq. \eqref{eq:TOAexpec} diverges but we can assign meaningful numerical results by interpreting the series as an asymptotic expansion of Eq. \eqref{eq:tau}. Furthermore, the closed form expression of Eq. \eqref{eq:TOAexpec} may be obtained by Borel resummation \cite{bender2013advanced,balser1994divergent}. 
	
	The asymptotic relation Eq. \eqref{eq:TOAexpec} shows that the leading term of the first infinite series corresponds to the relativistic time of arrival, while the rest are quantum correction terms. These terms arise due to the dispersion of the wavepacket $\varphi(q)$ as it propagates to arrive at $q=0$, which implies that the particle may be either delayed or advanced depending on the initial wavepacket \cite{galapon2009quantum}.  However, the effect of these quantum correction terms may be minimized up to an arbitrary order of $\hbar$ by imprinting an appropriate position-dependent phase on the initial wavefunction \cite{flores2016synchronizing}.  
	
	For a single peaked wavepacket $\varphi(q)$ centered around $q=q_o$, the expected relativistic TOA emerges from the first term of the first infinite series in Eq. \eqref{eq:TOAexpec}, explicitly we have
	\begin{align}
		t = -\dfrac{\mu q_o}{p} \left(  1 + \dfrac{2}{\pi} \int_1^\infty dz \dfrac{p^2}{(p^2+\mu^2 c^2 z^2)} \dfrac{\sqrt{z^2-1}}{z} \right).
		\label{eq:tclass}
	\end{align}
	The integral term  simplifies to 
	\begin{equation}
		\dfrac{2}{\pi} \int_1^\infty dz \dfrac{p^2}{(p^2+\mu^2 c^2 z^2)} \dfrac{\sqrt{z^2-1}}{z} = -1 + \sqrt{1+\dfrac{p^2}{\mu^2c^2}},
	\end{equation}
	which reduces Eq. \eqref{eq:tclass} to 
	\begin{equation}
		t = -\dfrac{\mu q_0}{p} \sqrt{1+\dfrac{p^2}{\mu^2c^2}}.
		\label{eq:tclassa}
	\end{equation}
	The quantum correction terms vanish as the momentum increases  since the factor $\gamma_c^{(n)}/p^n$ in Eq. \eqref{eq:TOAexpec} goes to zero. The only non-vanishing term is the relativistic time of arrival Eq. \eqref{eq:tclassa} which goes to $t=-q_o/c$. Thus, the expected TOA of a relativistic free particle is bounded by the TOA of a photon.

	\section{Quantum correction for Gaussian wavepackets}
	\label{sec:quantcorr}
	
	\begin{figure}[t]
		\centering
		\includegraphics[width=0.45\textwidth]{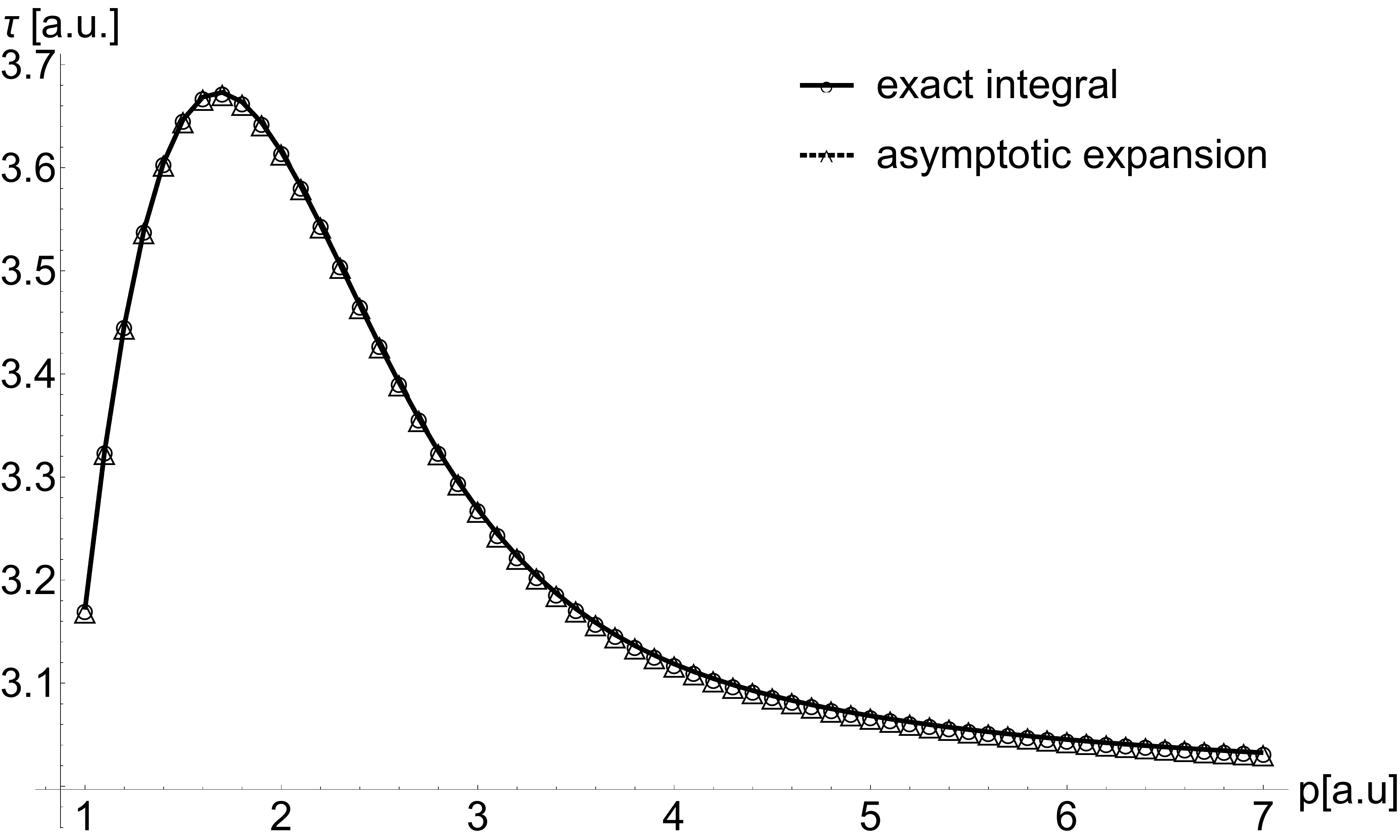}
		\caption{Comparison of the expectation value of the relativistic TOA operator given by the exact integral with the asymptotic expansion as the momentum increases with $q_o=-3$ and and $\sigma=0.5$ for the parameters $\mu=c=\hbar=1$.}
		\label{fig:compare}
	\end{figure}
	
	Let us now demonstrate the extent of these quantum correction terms by considering a single particle described as a Gaussian wavepacket 
	\begin{equation}
		\varphi(q) = \dfrac{1}{\sqrt{\sigma\sqrt{2\pi}}} \exp(-\dfrac{(q-q_0)^2}{4\sigma^2}). 
		\label{eq:gaussian}
	\end{equation}
	Since $\varphi(q)$ is a real valued function, then $\chi_2^{(n)}$ vanishes for all $n$ while 
	\begin{equation}
		\chi_1^{(n)} = q_0\dfrac{\Gamma(n+\frac{1}{2})}{\sqrt{\pi}(2\sigma^2)^n}.
	\end{equation}
	It follows from the asymptotic series Eq. \eqref{eq:TOAexpec} that the expected quantum time of arrival for a Gaussian wavepacket can be written as  $\tau=t Q_c(\mu,p,\sigma)$ wherein $Q_c(\mu,p,\sigma)$ is the quantum correction factor to the relativistic time of arrival, i.e. 
	\begin{align}
		Q_c = \left(1+\dfrac{p^2}{\mu^2c^2}\right)^{-1/2} \sum_{n=0}^{\infty} \dfrac{\hbar^{2n}}{p^{2n}} \dfrac{1}{(2\sigma^2)^n} \dfrac{\Gamma(n+\frac{1}{2})}{\sqrt{\pi}} \gamma_c^{(2n)}(p).
		\label{eq:qseries}
	\end{align}
	
	\begin{figure}[htp!]
		\centering
		\begin{subfigure}
			\centering
			\includegraphics[width=0.45\textwidth]{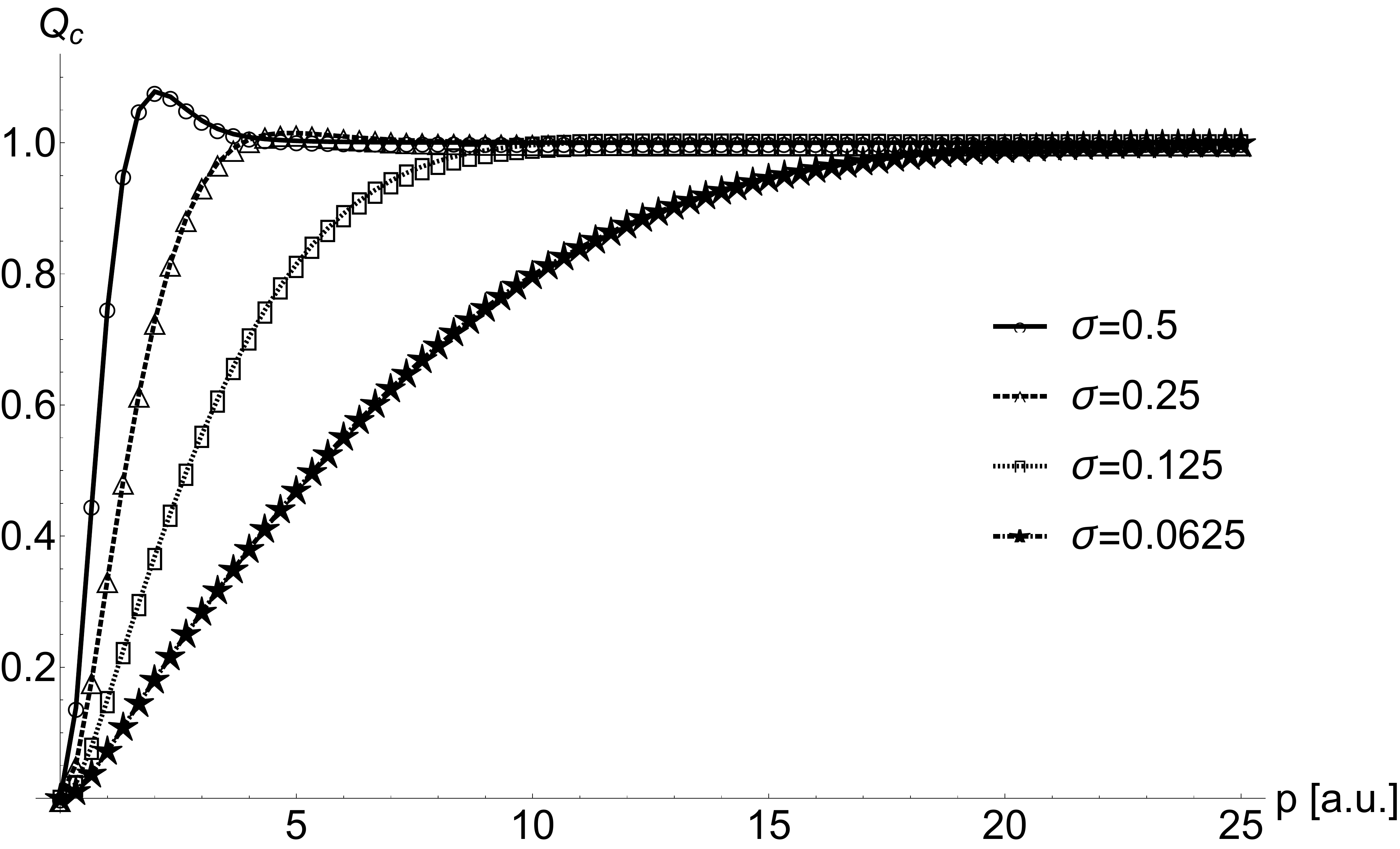}
		\end{subfigure} \hfil
		\begin{subfigure}
			\centering
			\includegraphics[width=0.45\textwidth]{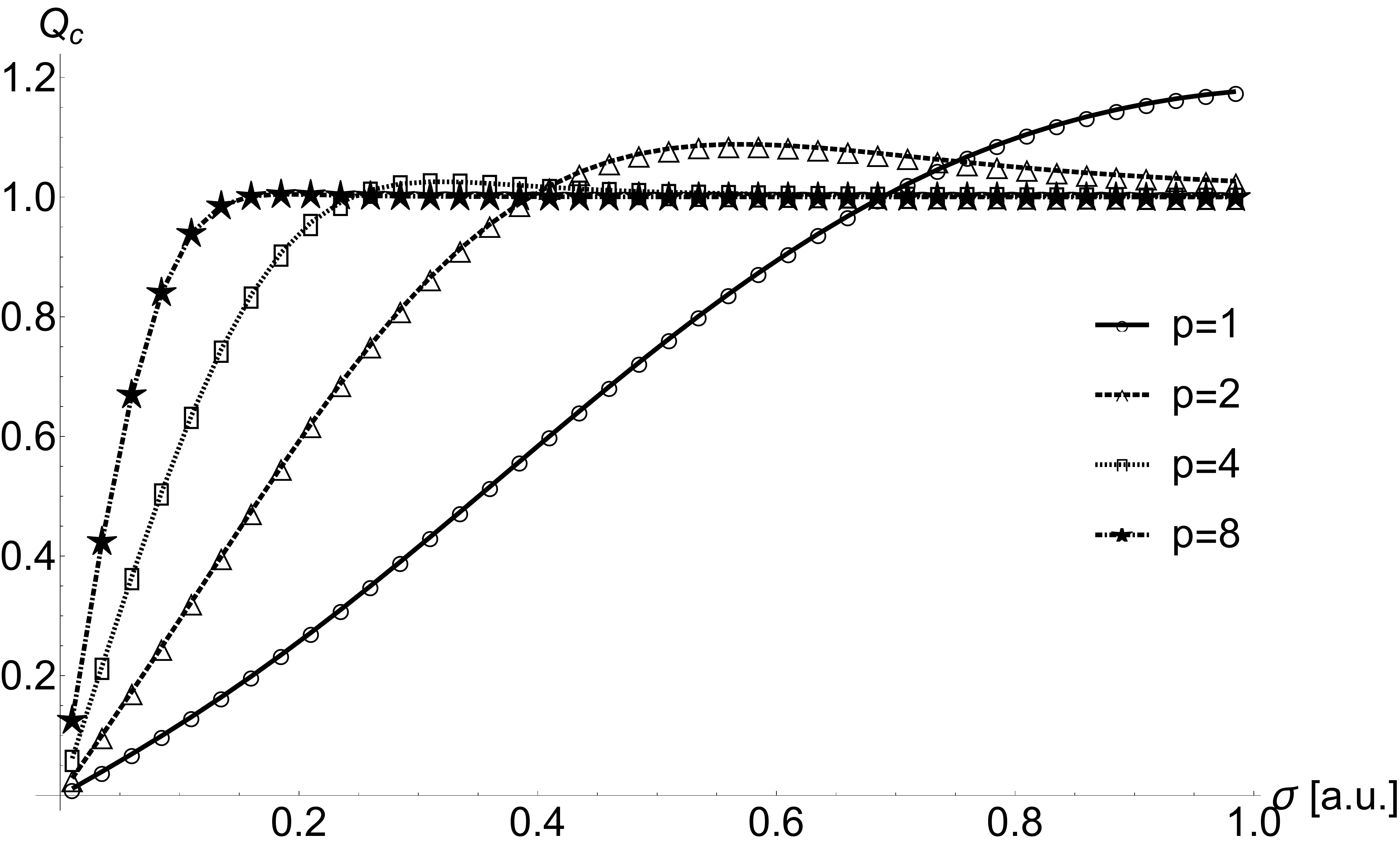}
		\end{subfigure} \hfil
		\caption{Effect of $\sigma$ and $p$ on the quantum correction factor for the parameters $\mu=c=\hbar=1$.}
		\label{fig:qfactor}
	\end{figure}	
	
	The series Eq. \eqref{eq:qseries} diverges because of $\Gamma(n+\frac{1}{2})$ but we can obtain a meaningful numerical result by performing a Borel ressumation. To do so, we replace the gamma function with its integral representation
	\begin{equation}
		\Gamma(n+\frac{1}{2}) = \int_0^\infty ds  e^{-s} s^{n-\frac{1}{2}},
	\end{equation}
	and interchange the orders of summation and integration which yields
	\begin{align}
		Q_c = \left(1+\dfrac{p^2}{\mu^2c^2}\right)^{-1/2} \left( Q_c^{(1)}+Q_c^{(2)}\right)
		\label{eq:Qc}
	\end{align}
	where, 
	\begin{align}
		Q_c^{(1)} = \dfrac{1}{\sqrt{\pi}} \text{P.V.} \int_0^\infty e^{-s} s^{-1/2} \left(1-\dfrac{\hbar^2}{p^2}\dfrac{s}{2\sigma^2}\right)^{-1}
		\label{eq:Qc1}
	\end{align}
	\begin{align}
		Q_c^{(2)} =&\dfrac{2}{\pi^{3/2}} \int_1^\infty dz \dfrac{\sqrt{z^2-1}}{z} \int_0^\infty ds e^{-s} s^{-1/2} \nonumber \\
		&\times \text{Re} \left[\dfrac{1}{1-i\frac{\mu c}{p}z} \left(1-\dfrac{1}{(1-i\frac{\mu c}{p}z)^2}\dfrac{\hbar^2}{p^2}\dfrac{s}{2\sigma^2}\right)^{-1} \right]
		\label{eq:Qc2}
	\end{align}

	For completeness, a comparison of the TOA  expectation value given by the exact integral Eqs. \eqref{eq:tau}-\eqref{eq:tauc}, and the asymptotic expansion leading to Eqs. \eqref{eq:Qc}-\eqref{eq:Qc2} is shown in Fig. \ref{fig:compare}. It can be seen that the two values are numerically equivalent. The effect of the quantum correction $Q_c$ is in shown Fig. \ref{fig:qfactor}. The TOA of a Gaussian wavepacket may be delayed ($Q_c>1$) or advanced ($Q_c<1$) but the effect of the quantum correction vanishes as $\sigma$ and $p$ increases.

	\section{Time of arrival distribution of Gaussian wavepackets}
	\label{sec:TOAdist}

\begin{figure}[t]
		\centering
		\includegraphics[width=0.4\textwidth]{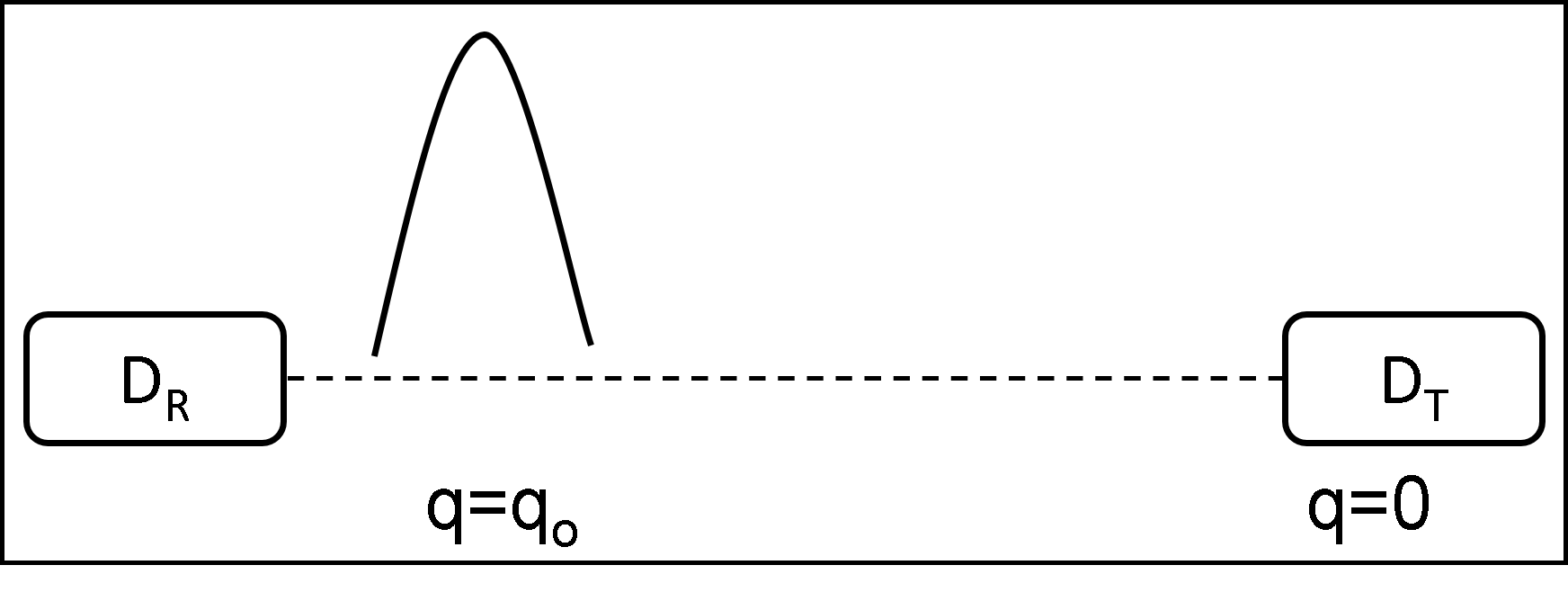}
		\caption{Time of arrival measurement scheme.}
		\label{fig:scheme}
	\end{figure}

	An ensemble of quantum particles prepared in the same initial state will not arrive at the origin at the same time. Instead, we get a time of arrival distribution that should peak at the expected time of arrival. Consider the measurement scheme shown in Fig. \ref{fig:scheme} as prescribed in Refs. \cite{pablico2020quantum} and \cite{galapon2012only} to provide an indirect and realistic way of obtaining the TOA of the particle. A wavepacket is initially prepared in a state $\psi(q)$ between two detectors $D_R$ and $D_T$. The detector $D_T$ is located at the arrival point $q=0$ and records the TOA whenever a particle passes through it, while the detector $D_R$ is placed in the far left of the wavepacket's initial position and does not record any data. After performing repeated measurements of identically prepared wavepakets, we then get a TOA distribution at $D_T$. 
	
	The TOA distribution is assumed to be generated by the spectral resolution of the operator $\mathsf{\hat{T}_{Ra}}$ using the non-nodal and nodal eigenfunctions. It was recently shown in Ref. \cite{sombillo2016particle} that for the non-relativistic free TOA operator, the non-nodal (nodal) eigenfunction is associated to particle arrival with detection (non-detection). We postulate that the same is true for the relativistic case, and only use the non-nodal eigenfunction to construct TOA distributions.

	For a particle prepared in an initial state $\ket{\psi}$, the probability that it will arrive at the origin, at a time t before $\tau$ is given by 
	\begin{equation}
		\mel{\psi}{\mathsf{\hat{\Pi}}}{\psi} = \int_{-\infty}^\tau dt \braket{\psi}{t} \braket{t}{\psi}
		\label{eq:povm}
	\end{equation}
	where $\mathsf{\hat{\Pi}}$ is a positive operator valued measure (POVM), and $\ket{\tau}$ is an eigenvector of the TOA operator. The TOA distribution is then constructed by differentiating Eq. \eqref{eq:povm} with respect to $\tau$ which yields
	\begin{align}
		\Pi_\psi(\tau) = \dfrac{d}{d\tau} \mel{\psi}{\mathsf{\hat{\Pi}}}{\psi} = \abs{\int_{-\infty}^{\infty} \psi^*(q) \phi_\tau(q)}^2
		\label{eq:toadist}
	\end{align}  
	where $\phi_\tau(q)$ is the TOA eigenfunction. It is easy to see that the TOA distribution is constructed by taking the overlap of the initial state with the TOA eigenfunction. The distribution Eq. \eqref{eq:toadist} represents the ideal distribution of a TOA experiment. In general, the measured TOA distribution will be deformed in the presence of a measuring instrument, and will be dependent on the details of the measuring instrument. 
	
	\begin{figure}[t!]
		\centering
		\subfigure[Coarse-grained non-nodal eigenfunction \label{fig:TOAdisttails-integ}]{\includegraphics[width=0.45\textwidth]{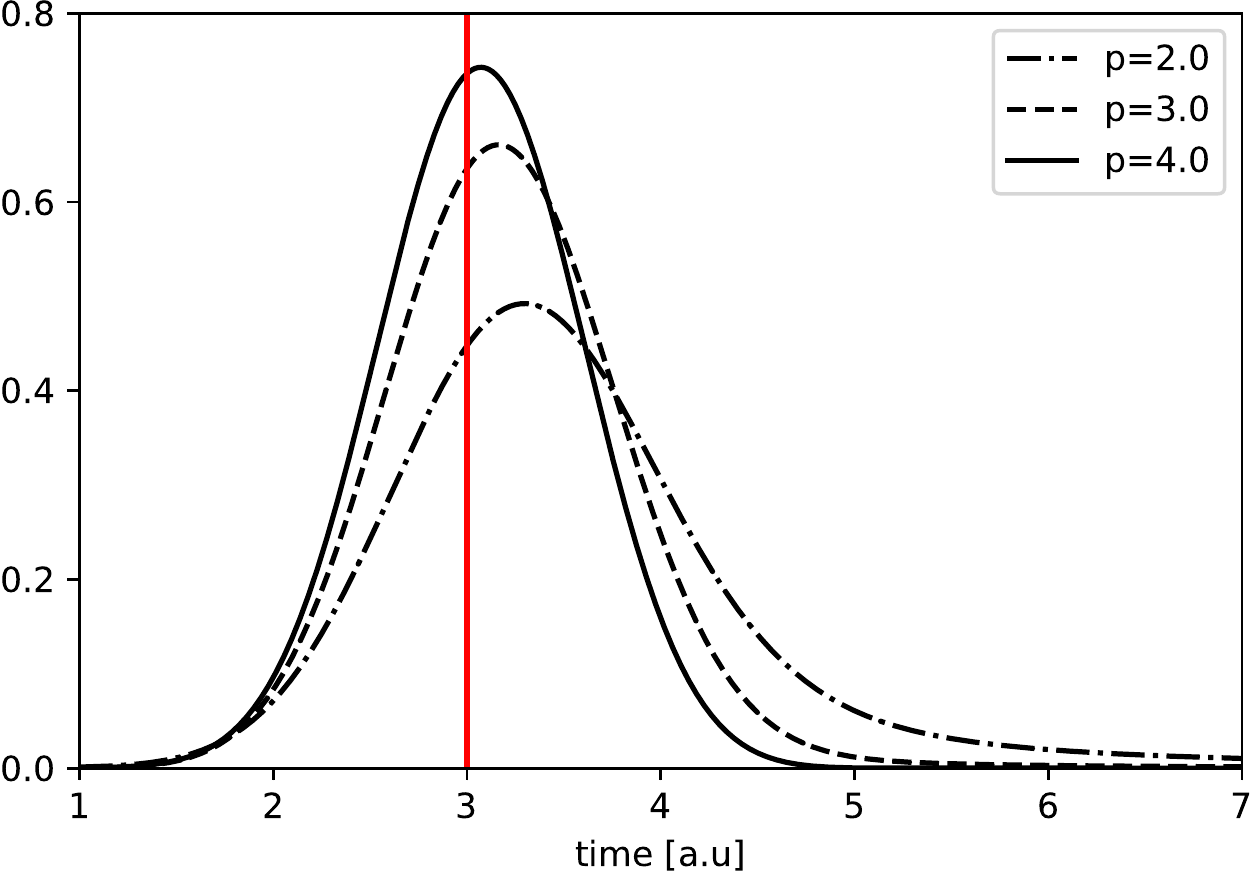}}\quad
		\subfigure[Analytic non-nodal eigenfunction\label{fig:TOAdisttails}]{\includegraphics[width=0.45\textwidth]{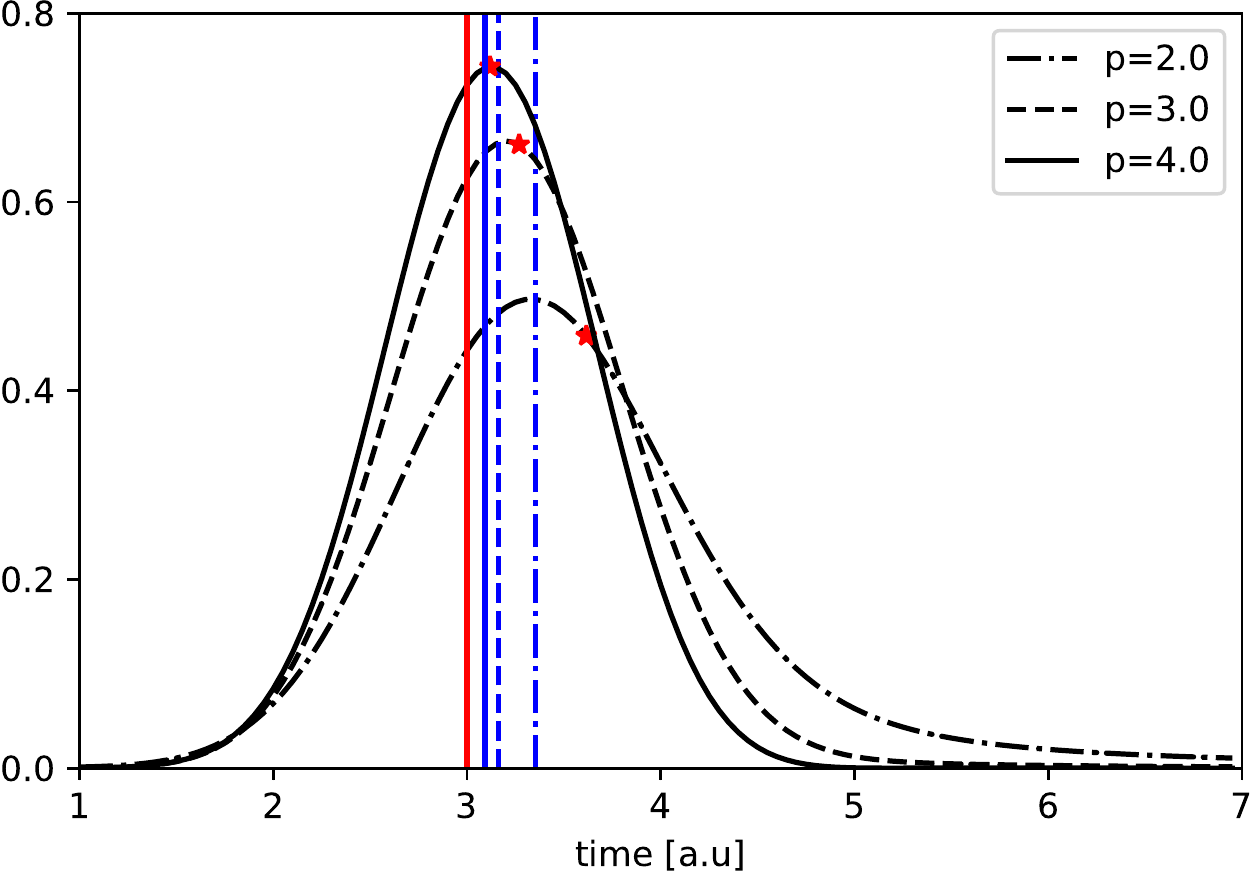}}
		\caption{Time of arrival distribution of a Gaussian wavepacket with initial position $q=-3$ and $\sigma=0.5$ for the parameters $\mu=\hbar=c=1$.  The blue line represents the relativistic time of arrival $t$, the red dots represent the expectation value $\tau=t Q_c$, and the red line represent the TOA of a photon.}
		\label{fig:TOAdist}
	\end{figure}
	
	The TOA distribution can be constructed by either using the non-nodal eigenfunction $\tilde{\Phi}_\tau(q)$ of the time kernel $\mel{q}{\mathsf{\hat{T}_{Ra}}}{q'}$, i.e.
	\begin{equation}
		\Pi_\psi(\tau) = \abs{\int_{-\infty}^\infty dq \psi^*(q) \tilde{\Phi}_\tau(q)}^2,
		\label{eq:toadistcoarse}
	\end{equation}
	or by using the the non-nodal eigeinfunctions Eq. \eqref{eq:noneig} from $\mathsf{\hat{T}_{Ra}}$ so that 
	\begin{equation}
		\Pi_\psi(\tau) = \abs{\int_{-\infty}^\infty dp \tilde{\psi}^*(p) \Phi_\tau^{(non)}(p)}^2,
		\label{eq:toadistanalytic}
	\end{equation}
	where, $\tilde{\psi}(p)$ is the Fourier transform of the initial state. 
	
\begin{figure}[t]
		\centering
		\includegraphics[width=0.45\textwidth]{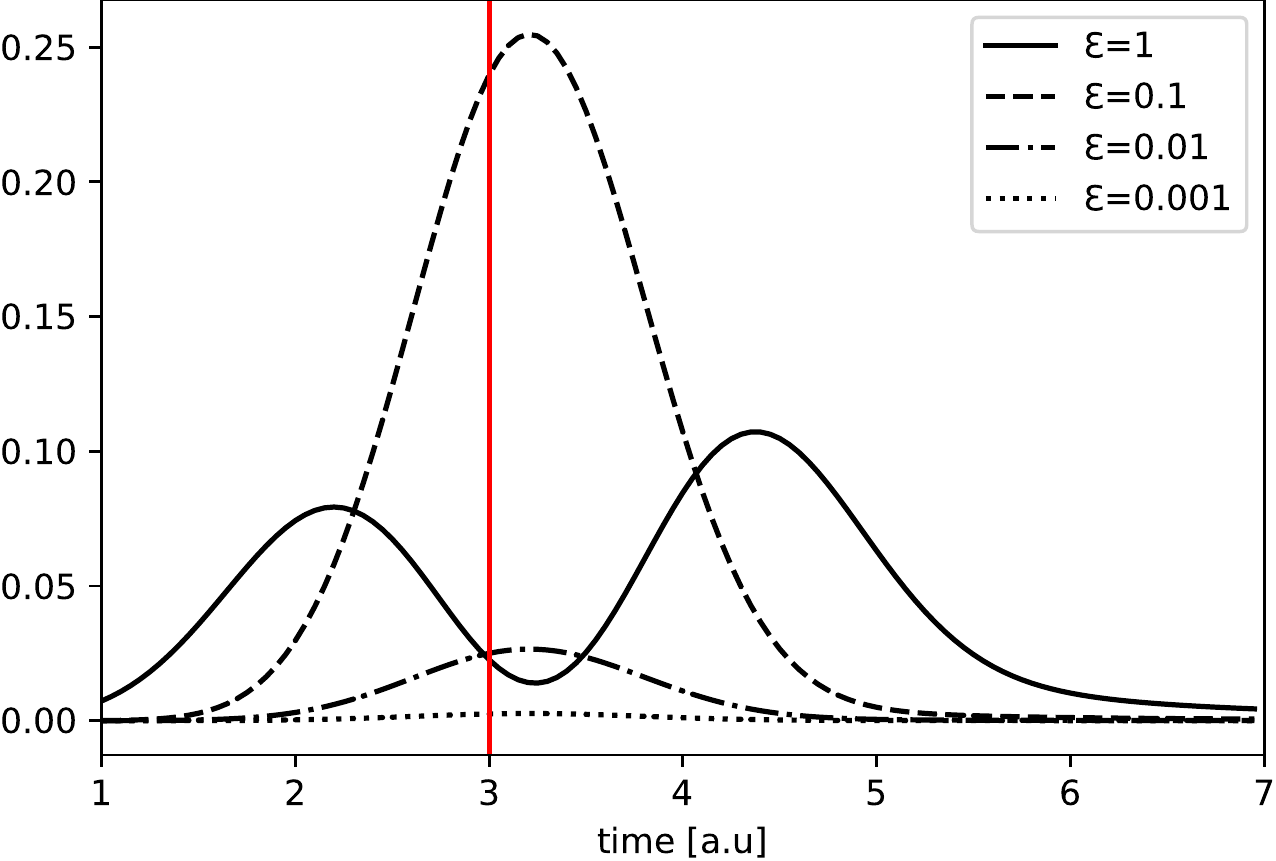}
		\caption{Time of arrival distribution using $\phi_{\tau}^{(Re)}(\epsilon,p)$ for a Gaussian wavepacket with initial position $q=-3$, momentum $p_o=3$, and $\sigma=0.5$ for the parameters $\mu=\hbar=c=1$ as $\epsilon\rightarrow0$. The red line represent the TOA of a photon.}
		\label{fig:TOAdistrazavi}
	\end{figure}

	Figure \ref{fig:TOAdisttails-integ} shows the constructed TOA distribution at the origin for a Gaussian wavepacket using Eq. \eqref{eq:toadistcoarse} via the coarse-grained TOA eigenfunctions of $\mel{q}{\mathsf{\hat{T}_{Ra}}}{q'}$ confined from $[-10,10]$ (see Appendix \ref{sec:coarsegrain} for details). Meanwhile, Figure \ref{fig:TOAdisttails} is the TOA distribution using Eq. \eqref{eq:toadistanalytic} via the non-nodal eigenfunctions Eq. \eqref{eq:noneig}. The same behavior is observed for both cases as expected. For completeness, a TOA distribution using Razavi's real-valued TOA eigenfunction $\phi_{\tau}^{(Re)}(\epsilon,p)$ is shown in Fig. \ref{fig:TOAdistrazavi}. It can be seen that as $\epsilon\rightarrow0$, the TOA distribution flattens out into a line which is not able to provide a meaningful interpretation for the distribution. This then means that the appropriate choice of TOA eigenfunctions that can give more insight on the quantum TOA problem is obtained by using the non-nodal and nodal eigenfunctions which were first obtained by taking the rigged Hilbert space extension of the operator $\mathsf{\hat{T}_{Ra}}$.

	The TOA distributions shown in Fig. \ref{fig:TOAdist} exhibits delayed arrival as the quantum expectation value $\tau$ is greater than the relativistic TOA $t$. Furthermore, it can be seen that the expectation value $\tau$ approaches $t$ since the quantum correction factor $Q_c$ approaches unity as the momentum increases. Lastly, the TOA distribution becomes sharper as the momentum increases which suggests that the particle becomes more \lq\lq classical\rq\rq. It is important to note that the peak of the TOA distributions, and expectation values $\tau$ in Fig. \ref{fig:TOAdist} are always to the right side of the TOA for a photon $t_{\text{photon}} =3$. This means that massive spin-0 particles will, on average, always arrive later than a photon. However, the TOA distribution can spread to values that are smaller than $t_{\text{photon}}$ which suggests that there is a non-zero probability for the particle to be \lq\lq superluminal\rq\rq but this behavior is not because the particle travels at a speed greater than the speed of light.   
	
	\begin{figure}[t!]
		\centering
		\subfigure[Coarse-grained non-nodal eigenfunction \label{fig:TOAdistcompact-integ}]{\includegraphics[width=0.45\textwidth]{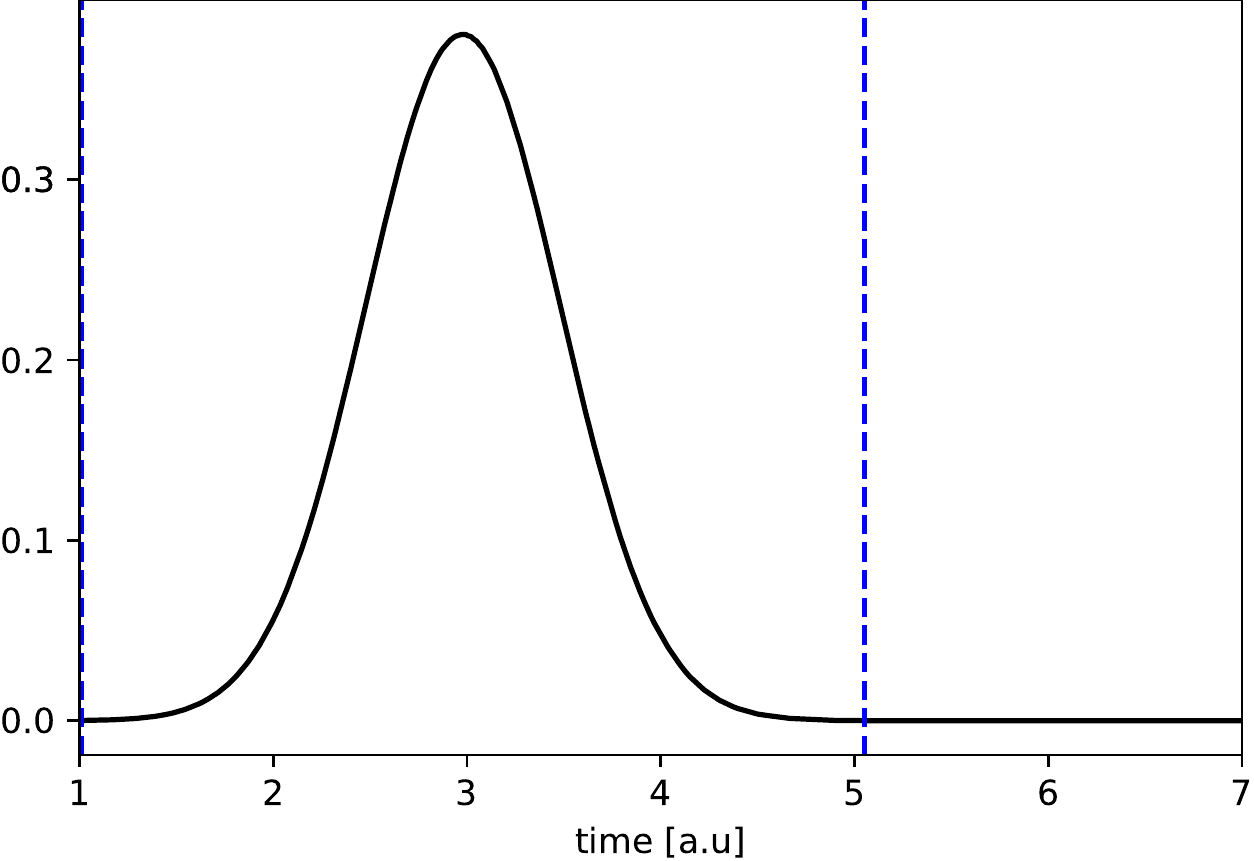}}\quad
		\subfigure[Analytic non-nodal eigenfunction\label{fig:TOAdistcompact}]{\includegraphics[width=0.45\textwidth]{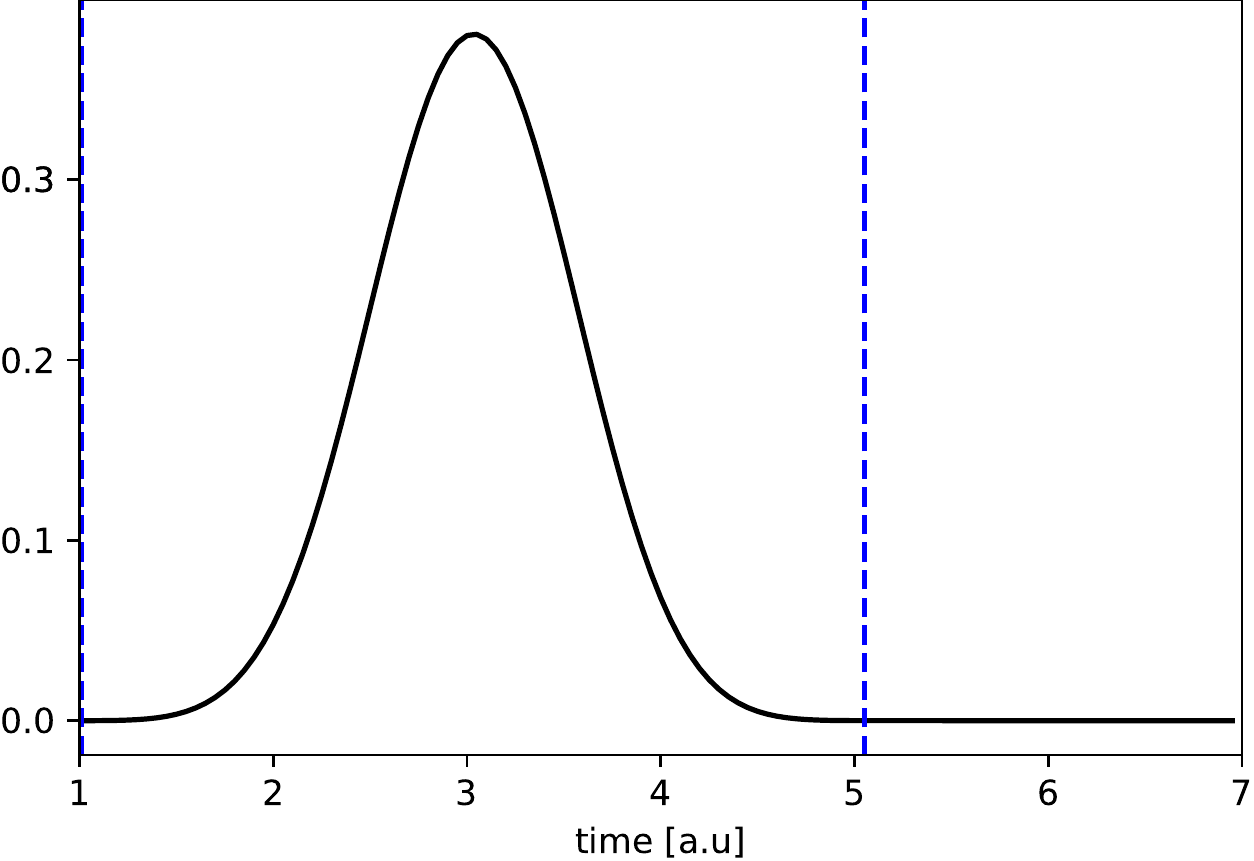}}
		\caption{Time of arrival distribution of a Gaussian wavepacket that has a compact support with initial position $q=-3$, momentum $p=7$ and $\sigma=0.5$ for the parameters $\mu=\hbar=c=1$.  The blue lines represent the relativistic TOA $t$ of a localized particle with initial position $q=-5$ and $q=-1$.}
		\label{fig:TOAdistcompact2}
	\end{figure}
	
	\begin{figure}[t!]
		\centering
		\subfigure[Position density distribution of the time-evolved state \label{fig:posdisttranslate}]{\includegraphics[width=0.45\textwidth]{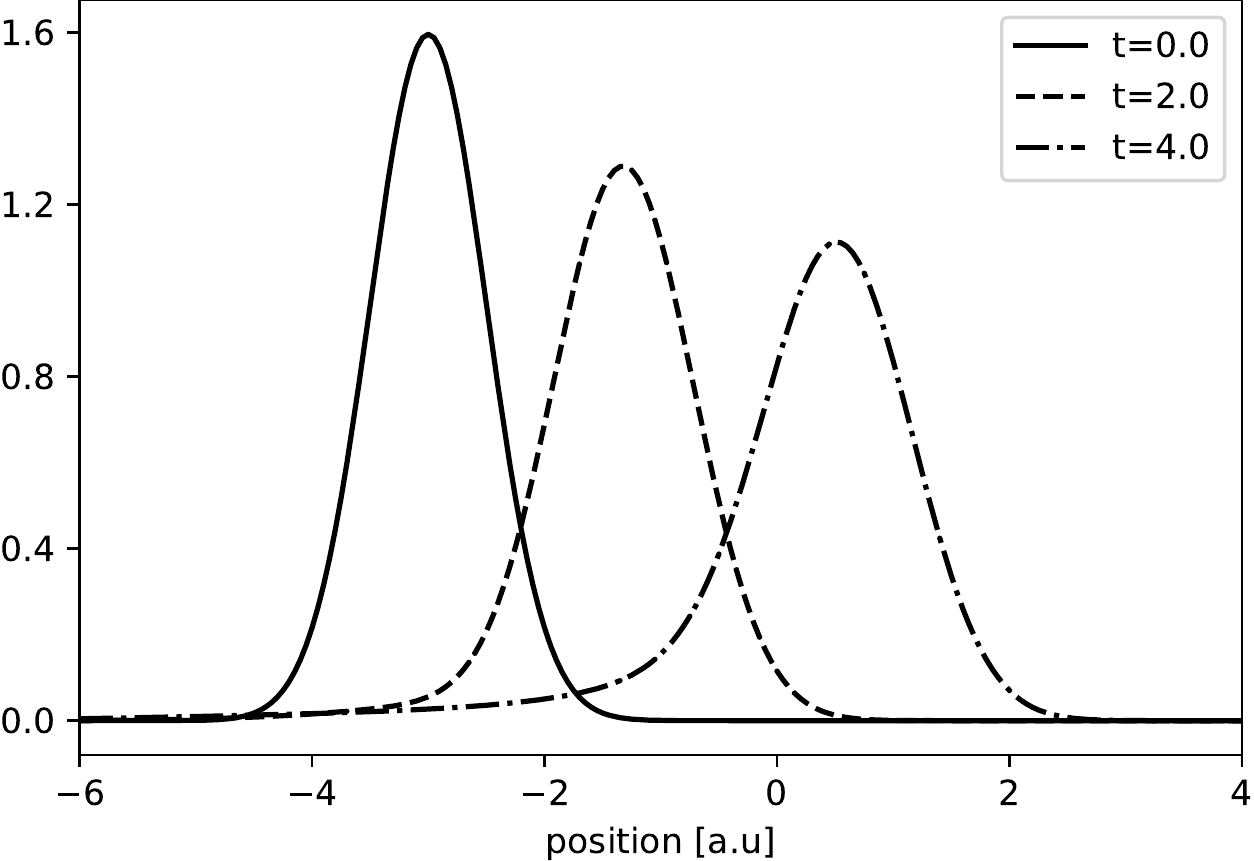}}\quad
		\subfigure[TOA density distribution of the time-evolved state\label{fig:toadisttranslate}]{\includegraphics[width=0.45\textwidth]{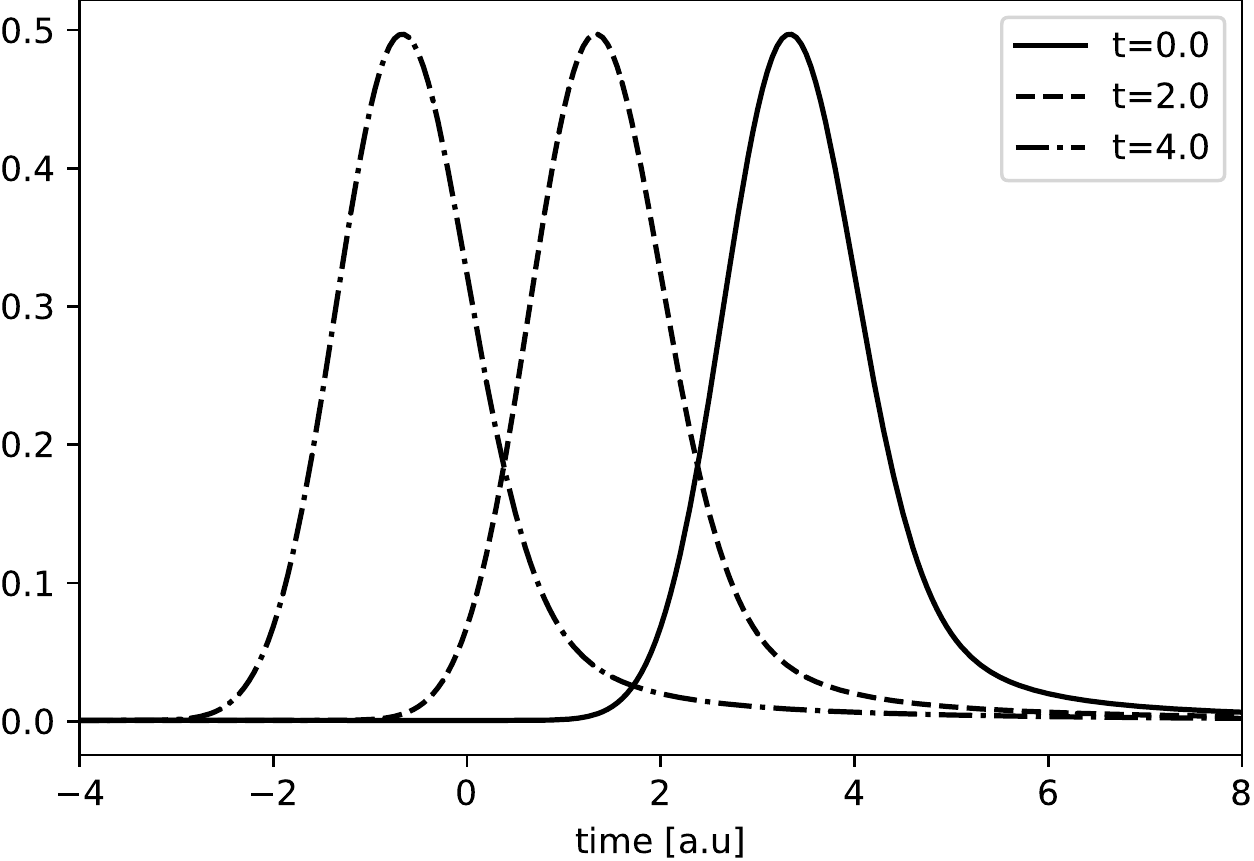}}
		\caption{Comparision of the position density distribution and TOA distribution of a time-evolved Gaussian wavepacket with initial position $q=-3$, momentum $p=2$ and $\sigma_q=0.5$ for the parameters $\mu=\hbar=c=1$.}
		\label{fig:disttranslate}
	\end{figure}
	
	The non-locality of the Gaussian wavepacket implies that the particle may be found in the neighborhood of its arrival point. This means that there is a non-zero probability that the particle may be found somewhere between its initial position and $q=0$ which causes the particle to arrive faster than $t_{\text{photon}} =3$ since it is nearer to the arrival point. To illustrate, consider an initial Gaussian wavepacket with compact support 
	\begin{equation}
		\psi_a(q) = \varphi(q) e^{ip_oq/\hbar}  \Theta((q-q_o)+a) \Theta(a-(q-q_o)).
	\end{equation}
	The support of $\psi_a(q)$ is chosen so that the tail of the Gaussian wavepacket does not extend to $\pm\infty$. Again, it can be seen in Fig. \ref{fig:TOAdistcompact2} that the same results are obtained using the coarse grained eigenfunctions of $\mel{q}{\mathsf{\hat{T}_{Ra}}}{q'}$ and the non-nodal eigenfunctions Eq. \eqref{eq:noneig}, which further supports our claim that the non-nodal and nodal TOA eigenfunctions provides more insight onto the quantum TOA problem. It is easy to see that the TOA distribution is now bounded by $\tau=tQ_c=1.011$ and $\tau=tQ_c=5.054$ which are the TOA of a localized particle with initial position $q=-1$ and $q=-5$, respectively. Thus, the \lq\lq superluminal\rq\rq TOA does not violate special relativity in the sense that the particle does not travel faster than the speed of light but because of its non-locality. 
	
	Lastly, having established that the coarse grained eigenfunctions of $\mel{q}{\mathsf{\hat{T}_{Ra}}}{q'}$ and the non-nodal eigenfunctions Eq. \eqref{eq:noneig} yields the same result, it will be convenient to use Eq. \eqref{eq:noneig} to show that the Hamiltonian and TOA operator are generators of translation of each other. This follows from the fact that $\mathsf{\hat{T}_{Ra}}$ is canonically conjugate to the free Hamiltonian  $\sqrt{\mathsf{\hat{p}}^2 c^2 + \mu^2 c^4}$. Specifically, if $\Pi_{\tilde{\psi}_o}(\tau)$ represents the TOA distribution of an initial state $\tilde{\psi}_o = \tilde{\psi}(p,t=0)$, then the TOA distribution for the time-evolved state $\tilde{\psi}(p,t) = \mathsf{\hat{U}_t}\tilde{\psi}(p,t=0) = e^{-iE_pt/\hbar}\tilde{\psi}(p,t=0)$is given as
	\begin{equation}
		\Pi_{\tilde{\psi}}(\tau - t) = \abs{\int_{-\infty}^\infty dp \tilde{\psi}^*(p,t) \Phi_\tau(p)}^2
	\end{equation} 
	It can be seen from Fig. \ref{fig:disttranslate} that the TOA distribution for the correspoding time-evolved state are just time translations of each other.

	\section{Summary and conclusion}
	\label{sec:conc}

	The relativistic free TOA operator $\mathsf{\hat{T}_{Ra}}$ is Hermitian and canonically conjugate with the system Hamiltonian, which are the basic requirements for a TOA operator. The rigged Hilbert space extenstion of $\mathsf{\hat{T}_{Ra}}$ was then constructed and shown that the eigenfunctions are two-fold degenerate which are called non-nodal and nodal. These eigenfunctions exhibit unitary arrival, which is also present for all the quantizations of the free and interacting case for the non-relativistic TOA operators.
	
	The TOA expectation value for a single-peaked wavepacket localized around $q=q_o$ was also shown to be equal to the relativistic TOA plus quantum correction terms. These correction terms imply that the TOA of a relativistic quantum particle may either be delayed or advanced. Furthermore, the effect of these correction terms vanish in the large momentum limit and that the expectation value is bounded by the TOA of a photon $t_\text{photon}=-q_o/c$. Thus, massive spin-0 particles will, on average, always arrive later than a photon. 
	
	The constructed TOA distributions were shown to be consistent with special relativity. That is, the peak of the distribution is always at a time greater than $t_\text{photon}$. The spread of the TOA distribution to times less than $t_\text{photon}$ was also shown to be a consequence of the non-locality of the wavepacket. Furthermore, the Hamiltonian was shown to be a generator of time translation which is a consequence of its conjugacy with the TOA operator. 
	
	These results give us confidence that it is possible to construct a meaningful relativistic TOA operator. Future studies may include the connection of a time-operator based theory of quantum TOA with the localization and appearance of a relativistic particle as an extension of the study done in Ref. \cite{galapon2009theory}. Moreover, a formalism for the quantization of a TOA operator for the interacting case may be useful in the tunneling time problem. That is, a formalism for the quantization of the relativistic TOA may provide new insights on whether the quantum particle becomes superluminal inside the barrier region, as current formalisms only use non-relativistic quantum mechanics.

	\section*{Acknowledgements}
	P.C.M. Flores acknowledges the support of the Department of Science and Technology – Science Education Institute through the ASTHRDP-NSC graduate scholarship program. This work was also supported by UP-System Enhanced Creative Work and Research Grant ECWRG 2019-05-R. 
	\appendix
	
	\section{Details on the calculation of the time kernel}
	\label{sec:pkernel}
	
	\begin{figure*}[t!]
		\centering
		\begin{subfigure}
			\centering
			\resizebox{0.35\textwidth}{!}{\begin{tikzpicture}

\def\gap{0.5}
\def\bigradius{5}
\def\littleradius{0.75}

\draw [help lines, -latex] (-1.25*\bigradius, 0) -- (1.25*\bigradius,0);
\draw [help lines, -latex] (0, -0.1*\bigradius) -- (0, 1.25*\bigradius);

\draw[line width=0.8pt, red, decorate, decoration=snake] (0, \bigradius/2) -- (0, 1.15*\bigradius);
\draw[thick] (0,0) node[branch point,draw=red,thick] {};
\draw[thick] (0,\bigradius/2) node[branch point,draw=red,thick] {};

\draw[line width=0.8pt,   decoration={ markings,
    mark=at position 0.13 with {\arrow[line width=1.2pt]{>}},
    mark=at position 0.26 with {\arrow[line width=1pt]{>}},
    mark=at position 0.35 with {\arrow[line width=1pt]{>}},
    mark=at position 0.443 with {\arrow[line width=1pt]{>}},
    mark=at position 0.575 with {\arrow[line width=1pt]{>}}},
    postaction={decorate}]
    let
        \n1 = {asin(\gap/2/\bigradius)},
        \n2 = {asin(\gap/2/\littleradius)}
    in (0:\bigradius) arc (0:90-\n1:\bigradius)
    -- (\gap/2, {\bigradius/2+\littleradius*cos(\n2)}) arc (90-\n2:-270+\n2:\littleradius)
    -- (90+\n1:\bigradius) arc (90+\n1:180:\bigradius)
    to [bend left = 15] (\bigradius, 0);
    
\draw[-latex, line width=0.8pt] (0, 0) -- ({\bigradius*cos(43)}, {\bigradius*sin(43)});
\draw[-latex, line width=0.8pt] (0, \bigradius/2) -- ({\littleradius*cos(-40)}, {\bigradius/2+\littleradius*sin(-40))});
\draw[-latex, line width=0.8pt] (0, \bigradius/2) -- ({\littleradius*cos(-40)}, {\bigradius/2+\littleradius*sin(-40))});

\node at ({1.125*\bigradius*cos(45)}, {1.125*\bigradius*sin(45)}) {$\gamma_{R1}^+$};
\node at (1.7*\gap, 0.83*\bigradius) {$\gamma_3^+$};
\node at (-1.6*\littleradius, \bigradius/2) {$\gamma_E^+$};
\node at (-1.7*\gap, 0.83*\bigradius) {$\gamma_4^+$};
\node at ({1.125*\bigradius*cos(135)}, {1.125*\bigradius*sin(135)}) {$\gamma_{R2}^+$};
\node at (0.5*\bigradius, 0.19*\bigradius) {$\gamma^+$};
\node at (0.5*\bigradius, -0.19*\bigradius) {};

\node at ({0.6*\bigradius*cos(51)}, {0.6*\bigradius*sin(51)}) {$R$};
\node at (\littleradius/2, 0.52*\bigradius) {$\epsilon$};

\end{tikzpicture}}
		\end{subfigure} 
		\begin{subfigure}
			\centering
			\resizebox{0.35\textwidth}{!}{\begin{tikzpicture}

\def\gap{0.5}
\def\bigradius{5}
\def\littleradius{0.75}

\draw [help lines, -latex] (-1.25*\bigradius, 0) -- (1.25*\bigradius,0);
\draw [help lines, -latex] (0, -0.1*\bigradius) -- (0, 1.25*\bigradius);

\draw[line width=0.8pt, red, decorate, decoration=snake] (0, \bigradius/2) -- (0, 1.15*\bigradius);
\draw[thick] (0,0) node[branch point,draw=red,thick] {};
\draw[thick] (0,\bigradius/2) node[branch point,draw=red,thick] {};

\draw[line width=0.8pt,   decoration={ markings,
    mark=at position 0.13 with {\arrow[line width=1.2pt]{>}},
    mark=at position 0.26 with {\arrow[line width=1pt]{>}},
    mark=at position 0.35 with {\arrow[line width=1pt]{>}},
    mark=at position 0.443 with {\arrow[line width=1pt]{>}},
    mark=at position 0.575 with {\arrow[line width=1pt]{>}}},
    postaction={decorate}]
    let
        \n1 = {asin(\gap/2/\bigradius)},
        \n2 = {asin(\gap/2/\littleradius)}
    in (0:\bigradius) arc (0:90-\n1:\bigradius)
    -- (\gap/2, {\bigradius/2+\littleradius*cos(\n2)}) arc (90-\n2:-270+\n2:\littleradius)
    -- (90+\n1:\bigradius) arc (90+\n1:180:\bigradius)
    to [bend right = 15] (\bigradius, 0);
    
\draw[-latex, line width=0.8pt] (0, 0) -- ({\bigradius*cos(43)}, {\bigradius*sin(43)});
\draw[-latex, line width=0.8pt] (0, \bigradius/2) -- ({\littleradius*cos(-40)}, {\bigradius/2+\littleradius*sin(-40))});

\node at ({1.125*\bigradius*cos(45)}, {1.125*\bigradius*sin(45)}) {$\gamma_{R1}^+$};
\node at (1.7*\gap, 0.83*\bigradius) {$\gamma_3^+$};
\node at (-1.6*\littleradius, \bigradius/2) {$\gamma_E^+$};
\node at (-1.7*\gap, 0.83*\bigradius) {$\gamma_4^+$};
\node at ({1.125*\bigradius*cos(135)}, {1.125*\bigradius*sin(135)}) {$\gamma_{R2}^+$};
\node at (0.5*\bigradius, -0.19*\bigradius) {$\gamma^-$};

\node at ({0.6*\bigradius*cos(51)}, {0.6*\bigradius*sin(51)}) {$R$};
\node at (\littleradius/2, 0.52*\bigradius) {$\epsilon$};

\end{tikzpicture}}
		\end{subfigure} \hfil
		\caption{Contours of integration for $q-q'>0$. }
		\label{fig:contour_pos}
	\end{figure*}
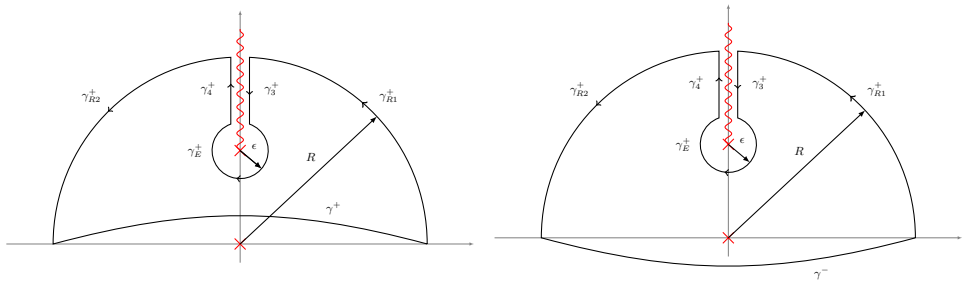
	
	The integral in Eq. \eqref{eq:pkernel0} has a pole at $p=0$, and branch points at $p=\pm i \mu c$. The Cauchy principal value is obtained by taking the average of the integral along the contours $\gamma^+$ and $\gamma^-$, that is,
	\begin{equation}
		\int_{-\infty}^{\infty} dp \dfrac{f(p)}{p} = \dfrac{1}{2} \left( \int_{\gamma^+} + \int_{\gamma^-}\right) \dfrac{f(z)}{z} dz 
		\label{eq:contour_pos}
	\end{equation}
	where, 
	\begin{align}
		f(p) =& \dfrac{1}{2\pi\hbar} \exp[\dfrac{i}{\hbar}(q-q')p] \frac{1}{p} \sqrt{1 + \frac{p^2}{\mu^2 c^2}}. 
	\end{align}
	To illustrate, we consider the case when $q-q'>0$ and evaluate Eq. \eqref{eq:contour_pos} using the contours shown in Fig. \ref{fig:contour_pos}. It is easy to show that as $\epsilon\rightarrow 0$ and $R\rightarrow\infty$, only the contribution of the branch cut and residue will remain. This now turns 
	\begin{align}
		\int_{-\infty}^{\infty}& \dfrac{dp}{2\pi\hbar} \exp[\dfrac{i}{\hbar}(q-q')p]\frac{1}{p} \sqrt{1 + \frac{p^2}{\mu^2 c^2}}  \nonumber \\
		=& \dfrac{i}{2\hbar}\Res[\exp[\dfrac{i}{\hbar}(q-q')z]\dfrac{1}{z}\sqrt{1+\dfrac{z^2}{\mu^2c^2}}]_{z=0} \nonumber \\
		&+\dfrac{i}{\hbar}  \dfrac{1}{\pi} \int_1^\infty dz \exp[-\dfrac{\mu c}{\hbar}(q-q')z]  \dfrac{\sqrt{z^2-1}}{z} \nonumber \\
		=& \dfrac{i}{2\hbar} \left( 1 + \dfrac{2}{\pi} \int_1^\infty dz \exp[-\dfrac{\mu c}{\hbar}(q-q')z]  \dfrac{\sqrt{z^2-1}}{z} \right)
		\label{eq:kernalevalodd}
	\end{align}

	The same process can be done when $q-q'<0$ wherein the contour will have to avoid to branch cut at $z=-i\mu c$ and the contour is closed in the clockwise direction. Thus, 
	\begin{align}
		\int_{-\infty}^{\infty}& \dfrac{dp}{2\pi\hbar} \exp[\dfrac{i}{\hbar}(q-q')p]\frac{1}{p} \sqrt{1 + \frac{p^2}{\mu^2 c^2}}  \nonumber \\
		=& \dfrac{i}{2\hbar} \left( 1 + \dfrac{2}{\pi} \int_1^\infty dz \exp[-\dfrac{\mu c}{\hbar}\abs{q-q'}z]  \dfrac{\sqrt{z^2-1}}{z} \right) \nonumber \\
		& \times  \text{sgn}(q-q').
		\label{eq:pkernela}
	\end{align}

	\section{Coarsegraining of the TOA operator}
	\label{sec:coarsegrain}
	
	For completeness, we outline the methods used by Refs. \cite{galapon2006theory,galapon2018quantizations} to construct the evolution of the TOA eigenfunctions and Refs. \cite{galapon2009theory,muga2000arrival,galapon2005transition} for the TOA distribution. This is done by numerically solving the eigenvalue equation 
	\begin{align}
		\int_{-\infty}^\infty dq' \mel{q}{\mathsf{\hat{T}_{Ra}}}{q'} \tilde{\Phi}(q') = \tau \tilde{\Phi}(q)
	\end{align}
	by replacing the bounds $(-\infty,\infty)$ to $[-l,l]$ and replacing the integral with a representative weighted sum
	\begin{align}
		\tau \tilde{\Phi}(q_k) \approx \sum_{l=1}^{2n+1} w_l  \mel{q_k}{\mathsf{\hat{T}_{Ra}}}{q_l} \tilde{\Phi}(q_l)
		\label{eq:matrix}
	\end{align}
	where, $2n+1$ is the number of quadrature points in the integration range $q_k$. The weights and abscissas are assigned using Gauss-Legendre quadrature method. The eigenvectors of Eq. \eqref{eq:matrix} are then sorted out as either non-nodal or nodal based on the $n^{th}$ element of the eigenvector, i.e. if the $n^{th}$ element is zero, then it is sorted as nodal, otherwise, it is non-nodal. The evolution of the state is then done in discrete time steps using the time evolution operator $\mathsf{\hat{U}_t} = e^{-iE_pt/\hbar}$, where $E_p = \sqrt{p^2 c^2 + \mu^2 c^4}$. The behavior as $l\rightarrow\infty$ is investigated by successively increasing the confining length. 
	
	The time of arrival distribution Eq. \eqref{eq:toadistcoarse} is then constructed by taking the overlap of the non-nodal eigenfunctions solved previously, with the initial state using Gauss-Legendre quadrature. The values are then interpolated to create a smooth curve.

	
	\bibliography{reltoaopr.bib}
	
\end{document}